\documentclass[twocolumn,aps,prd,amsmath,amssymb,preprintnumbers,nofootinbib]{revtex4-1}

\usepackage{graphicx,caption}
\usepackage{tensor}
\usepackage{xcolor}
\usepackage{braket}

\usepackage{multirow}

\usepackage{comment}

\begin{document}

\setcounter{footnote}{0}

\title{Emergent field theory}
\author{Erick I.\ Duque}
\email{eqd5272@psu.edu}
\affiliation{Institute for Gravitation and the Cosmos,
The Pennsylvania State University, 104 Davey Lab, University Park,
PA 16802, USA}

\begin{abstract}
The uniqueness theorems for general relativity and Yang--Mills theories can be circumvented by dropping the ubiquitous, yet often implicit, assumption that physical fields, such as the spacetime metric, are fundamental.
The novel concept of emergent fields makes it possible to construct modified theories of gravity and forces where the spacetime metric and strength tensor fields emerge from a covariance analysis in the canonical formulation with nontrivial relations to the fundamental phase space and no additional degrees of freedom are required.
This is an example of a post-Einstein--Yang--Mills theory that implies new physics.
In particular, explicit realizations of the theory in symmetry-reduced systems have shown robust resolutions of the singularities that plague the classical theories in regions of extreme spacetime curvature, including nonsingular (SU(2)$\times$U(1)-charged) black holes with a cosmologial constant and collapsing solutions, as well as Gowdy and FLRW cosmologies.
Further applications include modifications in the spectrum of quasinormal modes and in the evaporation process of black holes, as well as relativistic formulations of long-range gravitational effects capable of modeling MOND as an alternative solution to the dark matter problem.
New results here include an extension of the spherically symmetric system that couples SU(2) gauge fields and the generalization of previous dynamical, homogeneous solutions.
\end{abstract}

\maketitle

\section{Introduction}

In the last decades, much of the effort in improving our understanding of gravity is motivated by two theoretical problems of general relativity: The singularities that plague it in its classical form and its incompatibility with quantum mechanics.
A long-awaited revolution is expected to quantize the gravitational field and it is often assumed that it will also resolve the classical singularities.
However, several technical and conceptual difficulties stand in the way towards the promised revolution, including the inadequacy of renormalization techniques in the perturbative quantization of general relativity (which introduces new singularities) and the role of measurements in the quantum theory.
Leading proposals have not overcome these challenges yet, often introducing additional problems of their own: Either requiring the introduction of extra dimensions, new (super)symmetries, and new particles or degrees of freedom that remain untestable hypotheses, or of regularization schemes that discretize space and correspondingly modify the Hamiltonian of general relativity, obscuring the preservation of the principle of general covariance.

The absence of resolutions the these long-standing challenges give rise to the question of whether we are missing something in our understanding of gravity prior to its quantization.
After all, there are several unexplained phenomena from observations at seemingly classical (certainly large) scales, including the \emph{dark matter} and the \emph{dark energy} problems.
Each of these problems is typically assumed to be a lacuna in our understanding of the matter content of the universe if systematics issues of experiments or observations can be discarded.
The traditional strategy towards these problems has been the continuation of a road that led twentieth-century physics to great success: Explain an anomalous observation by the introduction of a new particle.
This new-particles hypothesis goes beyond the standard model of particle physics, but usually still describes gravity with general relativity.
However, none of such proposals have been confirmed by direct detections of the corresponding new forms of matter after decades of much effort to do so.
If we reject the new-particles hypothesis to explain the astronomical and cosmological phenomena that challenge the established physical theories, then all of these are potential cases for modified gravity or matter dynamics without new degrees of freedom.

A difficulty in formulating modified gravity or matter dynamics lies in the fact that the action of the standard model and general relativity is strongly restricted by their symmetries and can even be derived from them, up to a small set of undetermined constants.
For instance, the Einstein--Yang--Mills action must be invariant under coordinate changes and SU($n$) transformations ---as well as under parity or time-reversal transformations as additional symmetries for the simpler cases.
Since the action is a functional of the spacetime metric $g_{\mu\nu}$ and the $\mathfrak{su}(n)$ connection $A^i_\mu$, it is given by curvature invariants.
In vacuum general relativity, the action is uniquely determined by the spacetime Ricci scalar and the invariant spacetime volume, up to the choice of Newton's and a cosmological constant, via Lovelock's theorem in four dimensions if it is restricted to depend on up to second-order derivatives of the metric \cite{Lovelock1,Lovelock2}.
Higher-order derivatives can be included in the action, for instance by replacing the Ricci scalar $R$ by a non-linear function $f(R)$ or by constructing invariants from contracting the components of multiple Riemann tensors.
However, such higher-curvature terms typically imply additional degrees of freedom that result in instabilities \cite{ostrogradsky1,ostrogradsky2,woodard}.
Similarly, the Yang--Mills sector is determined by the strength tensor of the connection contracted with itself (in both spacetime and SU($n$) indices) plus a topological $\theta$-term \cite{Weinberg}.

A similar uniqueness theorem for general relativity exists in the canonical formulation \cite{ADM,arnowitt2008republication,hojman1976geometrodynamics,kuchar1974geometrodynamics}.
However, due to the inherent slicing of the manifold into three-dimensional hypersurfaces and the distinction between configuration variables and their momenta, spacetime covariance is not manifest in the canonical formulation.
However, the dynamical equations, composed not only of Hamilton's equations of motion but also constraints, are equivalent to the Einstein equation and are therefore covariant.

In its canonical formulation, general relativity is a fully constrained theory \cite{Dirac,Katz,ADM,arnowitt2008republication,hojman1976geometrodynamics}: The Hamiltonian is a linear combination of first-class constraints.
Therefore, the constraints generate evolution equations as well as gauge transformations by their Hamiltonian vector fields.
On shell, when the constraints and equations of motion hold, these gauge transformations are equivalent to spacetime diffeomorphisms.

As shown recently \cite{EMG,EMGCov}, new theories of modified gravity are possible in a canonical formulation because some assumptions used to specify an action can be weakened.
In particular, the starting point of an action principle is the assumption of a fundamental spacetime metric with respect to which the action is minimized to generate Lagrange's equations of motion; furthermore, this same metric defines an invariant four-volume that is used as an integration measure in the action.
On the other hand, the starting point of the canonical formulation only requires spatial tensors with a phase-space structure to generate Hamilton's equations of motion; moreover, only spatial integrations are required for the Hamiltonian which do not even require a spatial metric for the integration measure because the momenta are inherently densitized spatial tensors.
In general relativity, the configuration variable of the phase space is identical to the three-dimensional spatial metric $q_{ab}$ induced by the spacetime metric on the initial hypersurface of the canonical analysis.
This identity, however, can be relaxed such that the spacetime metric is not fundamental but rather emergent.
The induced spatial metric $\tilde{q}_{ab}$ of the corresponding emergent spacetime geometry does not agree with a canonical configuration variable, but it is uniquely determined by the structure function of the Poisson bracket of two Hamiltonian constraints, which is given by the inverse of the spatial metric in general relativity.
This key concept of emergence is what gave the theory the name \emph{emergent modified gravity} \cite{EMG,EMGCov}.

Unlike in Lagrangian formulations with fundamental spacetime tensors, general covariance is cannot incorporated in the canonical formulation in such a straightforward manner; it must instead be demonstrated by use of Poisson brackets and gauge flows of the constraints.
In emergent modified gravity, covariance conditions are formulated such that the gauge transformations of the emergent spacetime metric $\tilde{g}_{\mu\nu}$ generated by the constraints correspond precisely to coordinate transformations.
These covariance conditions imply strong restrictions on the allowed modifications of the constraints with respect to general relativity and also determine the relation between the phase-space variables and the spacetime geometry.
However, replacing the assumption of the latter relation for its derivation turns out to give greater freedom in comparison to general relativity and allows the circumvention of uniqueness theorems that assume that the metric is a fundamental field such that modified gravity is possible without introducing new degrees of freedom.

Introducing Yang--Mills fields to general relativity in the canonical formulation implies additional first-class Gauss constraints which generate the corresponding SU($n$) transformations.
While the Poisson bracket between two Gauss constraints reflects the underlying $\mathfrak{su}(n)$ algebra with only structure \textit{constants}, the brackets of the full set of constraints forms an algebra with additional structure functions.
While the gravitational structure function is given by the inverse of the spatial metric, the Yang--Mills structure functions are given by the components of the strength tensor of the connection.
In the same spirit of emergent modified gravity, these new structure functions can instead define emergent force fields with nontrivial relations to the fundamental phase space.
This was foreseen in a previous study of electromagnetism \cite{EMGEM}, where the concept of an emergent electric field was introduced.
Here, we systematically generalize the concept of emergent fields to encompass the Yang--Mills forces.
We refer to this more general procedure as \emph{emergent field theory}.

Emergent field theory is not a quantum theory, but it is an attempt to go one step beyond the standard model and general relativity.
This step is the introduction of the emergence of physical fields from the gauge symmetries of covariant Hamiltonian systems, which in turn enables the formulation of modified gravity and matter dynamics without extra degrees of freedom.
Moreover, additional forms of matter can be consistently coupled, including scalar matter \cite{EMGscalar} and perfect fluids \cite{EMGPF}.
In particular, applications in spherical symmetry include nonsingular, charged black hole solutions \cite{Alonso_Bardaji_2022,ELBH,EMGEM}, as well as nonsingular solutions for the homogeneous collapse of spherical scalar matter \cite{EMGscalar}, generalizations of which we present here.
Furthermore, nonsingular solutions have also been obtained beyond spherically symmetric cases such as Gowdy systems \cite{EMGGowdy}, as well as in cosmological systems \cite{EMGCosmoK,EMGCosmo,EMGCosmoD}.

Since the central concept of emergent fields relies on the gauge symmetries in canonical form, we begin with a review of the canonical formulation of the classical Einstein--Yang--Mills system in Section~\ref{sec:Canonical formulation}.
The general procedure for emergent field theory is outlined in Section~\ref{sec:Emergent fields}.
Applications in spherical symmetry, including the new coupling of SU(2) gauge fields and a demonstration of nonsingular solutions, are presented in Section~\ref{sec:Spherically symmetric system}.
Finally, Section~\ref{sec:Discussion} contains a discussion of further applications of the theory, as well as the prospects for future developments and potential implications for the quantum theory.
In what follows, we work in natural units $c=G=\epsilon_0=1$ such that the dimensions of all quantities are suitable powers of length $L$.
Greek letters are used for spacetime indices, the Latin letters $a,b,c,\dots,h$ are used for spatial indices, and $i,j,k,\dots$ for SU($n$) indices.

\section{Canonical formulation}
\label{sec:Canonical formulation}

In the canonical formulation, the spacetime region of interest is assumed hyperbolic, $M = \Sigma \times \mathbb{R}$, with a 3-dimensional spatial manifold $\Sigma$.
Given a foliation of $M$ into spacelike hypersurfaces $\Sigma_t$ parametrized by $t\in{\mathbb R}$, the spacetime metric
$g_{\mu\nu}$ on $M$ defines the unit vector field $n^{\mu}$ normal to $\Sigma_t$, and induces the spatial metric $q_{ab}(t)$ on
$\Sigma_{t}$ by the restriction of the spacetime tensor $q_{\mu\nu}=g_{\mu\nu}+n_{\mu}n_{\nu}$ to $\Sigma_{t}$.
Unambiguous time-evolution between adjacent hypersurfaces is given by a time-evolution vector field
\begin{equation}
  t^\mu = N n^\mu + N^a s_a^\mu\,,
  \label{eq:Time-evolution vector field}
\end{equation}
where $s_a^\mu$ are basis vectors on the spatial hypersurfaces, $s_a^\mu(t): T\Sigma_{t}\to TM$, the normal component $N$ is the lapse function and the tangential component $N^a$ is the shift vector field which appear in the line element
\begin{equation}
  {\rm d} s^2 = - N^2 {\rm d} t^2 + q_{a b} ( {\rm d} x^a + N^a {\rm d} t )
  ( {\rm d} x^b + N^b {\rm d} t )
  \,.
  \label{eq:ADM line element}
\end{equation}
This canonical decomposition of the Einstein--Hilbert action results in the spatial metric components $q_{ab}$ being the configuration variables with conjugate momenta $p^{ab}=\frac{\sqrt{\det q}}{16\pi} \left(K^{ab}-K^c_c q^{ab}\right)$, where $K_{ab}$ is the extrinsic curvature of the hypersurface $\Sigma_t$, while $N$ and $N^a$ appear as Lagrange multipliers that imply the constraints $H^{\rm EH}$ and $H_a^{\rm EH}$ which must vanish on physical solutions:
\begin{eqnarray}
    S_{\rm EH} 
    \!\!&=&\!\! \frac{1}{16\pi}\int{\rm d}^4x\sqrt{-\det g} \left[R-2\Lambda\right]
    \nonumber\\
    \!\!&=&\!\! \int {\rm d}t {\rm d}^3x\left(p^{ab}\dot{q}_{ab}- H^{\rm EH}N - H^{\rm EH}_a N^a\right)\,,
\end{eqnarray}
with
\begin{eqnarray}
    H^{\rm EH} \!\!&=&\!\! - \frac{\sqrt{\det q}}{16 \pi} \left(\tensor[^{(3)}]{\!R}{}-2\Lambda\right)
    \nonumber\\
    \!\!&&\!\!
    + \frac{16\pi}{\sqrt{\det q}} \left(p_{ab}p^{ab} - \frac{\left(p^a_a\right)^2}{2}\right)\,,\label{eq:Hamiltonian constraint - full GR}
    \\
    H_a^{\rm EH} \!\!&=&\!\! - 2 D_b p^b_a
    \,,\label{eq:Vector constraint - full GR}
\end{eqnarray}
where $\tensor[^{(3)}]{\!R}{}$ and $D_b$ are, respectively, the Ricci scalar and the covariant derivative associated to the spatial metric $q_{ab}$.

\begin{figure}[!htb]
    \centering
    \includegraphics[trim=6cm 3cm 6cm 3cm,clip=true,width=1\columnwidth]{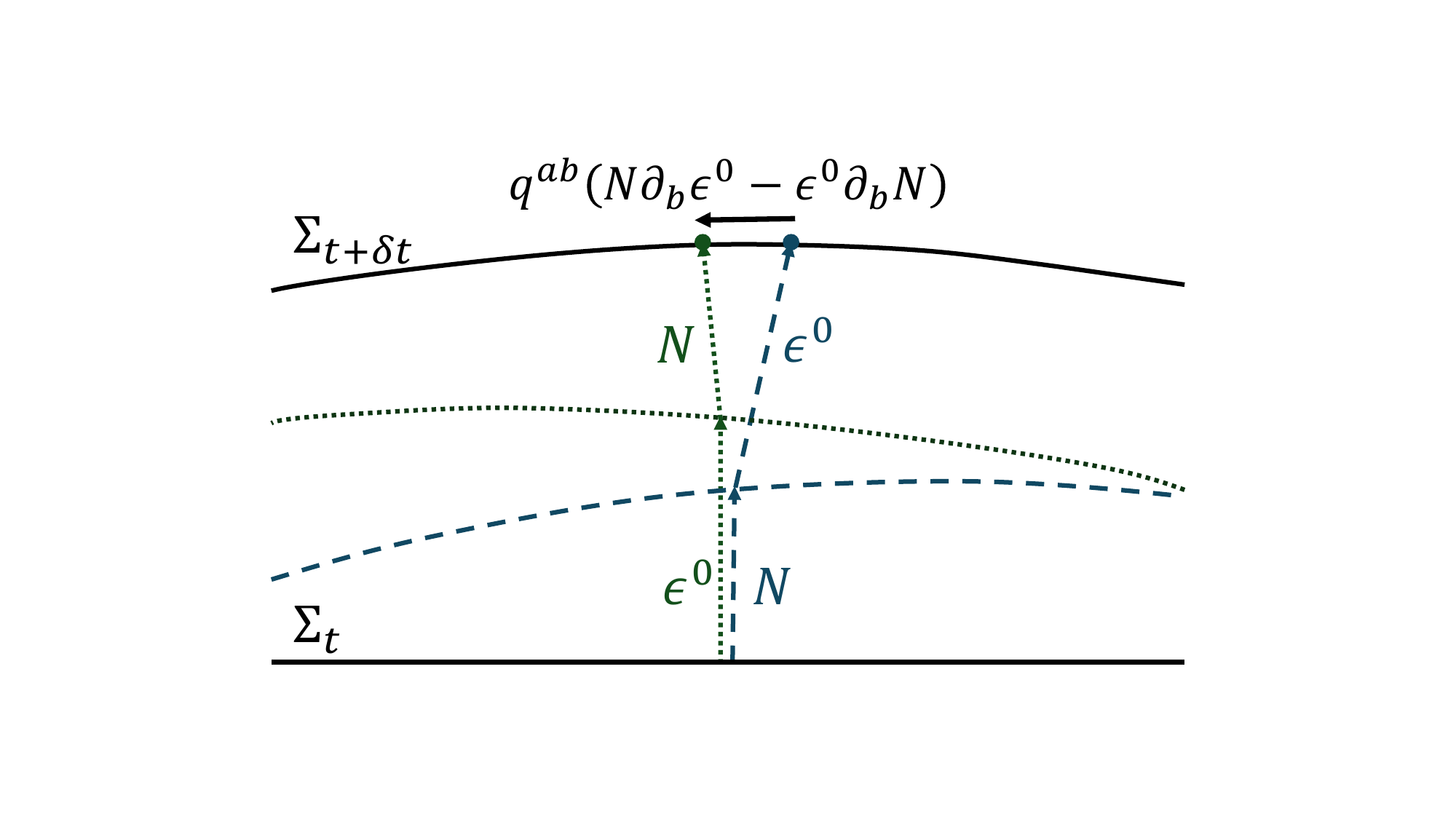}
    \caption{Two consecutive Lorentz-normal deformations of length $N$ and $\epsilon^0$ of a hypersurface $\Sigma_t$ take the same point from $\Sigma_t$ to different points in the deformed hypersurface $\Sigma_{t+\delta t}$ if the order of the deformations is reversed. The two final points are connected to each other by a deformation $q^{ab}(N\partial_b\epsilon^0-M\partial_b\epsilon^0)$ tangential to $\Sigma_{t+\delta t}$. This deformation commutator illustrates the geometric content of the bracket (\ref{eq:Hypersurface deformation algebra - HH}).}
    \label{fig:HDA}
\end{figure}

The constraints play a dual role as the time-evolution generator $H[N,N^a]=H^{\rm EH}[N]+H^{\rm EH}_a[N^a]$ (where $H^{\rm EH}[N]=\int {\rm d} x^3 H^{\rm EH}(x) N(x)$) when smeared by lapse and shift and as the gauge transformation generator $H[\epsilon^0,\epsilon^a]$ when smeared by arbitrary gauge functions $\epsilon^0$ and $\epsilon^a$:
For a phase-space function $\mathcal{O}$, its gauge transformation is given by $\delta_\epsilon \mathcal{O} = \{ \mathcal{O} , H^{\rm EH}[\epsilon^0 , \epsilon^a]
\}$ and its time-evolution by $\Dot{\mathcal{O}} = \{ \mathcal{O} , H^{\rm EH}[N , N^a] \}$.
The Poisson brackets of the constraints with themselves form a first-class algebra,
\begin{subequations}\label{eq:Hypersurface deformation algebra}
\begin{eqnarray}
    \{H^{\rm EH}_a[N^a],H^{\rm EH}_b[\epsilon^b]\}
    \!\!&=&\!\! - H^{\rm EH}_a\left[\mathcal{L}_{\vec{\epsilon}}N^a\right]
    \\
    \{H^{\rm EH}[N],H^{\rm EH}_a[\epsilon^a]\} \!\!&=&\!\! - H^{\rm EH}[\epsilon^a\partial_a N]
    \\
    \{H^{\rm EH}[N],H^{\rm EH}[\epsilon^0]\} \!\!&=&\!\! - H^{\rm EH}_a\left[q^{ab} \left(\epsilon^0\partial_bN-N\partial_b\epsilon^0\right)\right]\quad\quad\!\!\!
    \label{eq:Hypersurface deformation algebra - HH}
\end{eqnarray}
\end{subequations}
which implies that they are preserved on shell under arbitrary gauge transformations and time evolution.
The off-shell algebra describes the gauge content, closely related to the underlying diffeomorphism covariance, of general relativity in terms of hypersurface deformations, see Fig.~\ref{fig:HDA}.
Unlike typical Lie algebras, (\ref{eq:Hypersurface deformation algebra}) possesses structure \emph{functions} given by the components of the inverse of the spatial metric $q^{ab}$.

Similarly, the canonical decomposition of the Yang--Mills action results in the contributions $H^{\rm YM}$ and $H_a^{\rm YM}$ to the constraints $H=H^{\rm EH}+H^{\rm YM}$ and $H_a=H_a^{\rm EH}+H_a^{\rm YM}$ with the spatial components of the SU($n$) connection $A^i_a$ as configuration variables (where $i=1,2,\dotsi,n^2-1$) with conjugate momenta $P^a_i=\delta_{ij}\sqrt{\det q} \left( q^{a b} F_{0 b}^j - \frac{\theta}{2} \epsilon^{a b c} F_{b c}^j \right)=:-E^a_i-\theta B^a_i$ related to the components of the strength tensor field $F_{\mu\nu}^i=2\partial_{[\mu}A^i_{\nu]}+\tensor{f}{^i_j_k}A_\mu^jA^k_\nu$, where $F^i_{0a}=n^\mu F_{\mu a}^i$, $E^a_i$ is the electric field, $B^a_i$ is the magnetic field, and $\theta$ is the constant of the topological term, while the time components $A^i_t$ appear as Lagrange multipliers that imply the Gauss constraint $G_i$ in the Yang--Mills action contribution
\begin{eqnarray}
    \!\!S_{\rm YM} \!\!\!&=&\!\!\! - \frac{1}{4} \int{\rm} {\rm d}^4x \sqrt{- \det g}\left[F_i^{\mu\nu} F_{\mu\nu}^i - \frac{\theta}{2} \tensor{\epsilon}{^\mu^\nu_\alpha_\beta} F^i_{\mu \nu}F_i^{\alpha \beta}\right]
    \nonumber\\
    \!\!\!&=&\!\!\! \int{\rm} {\rm d}t{\rm d}^3x \left[P^a_i\dot{A}^i_a - H^{\rm YM}N -H^{\rm YM}_aN^a-G_iA^i_t\right]
    \!,
\end{eqnarray}
where
\begin{eqnarray}
    \label{eq:Hamiltonian constraint YM contribution - YM - classical}
    H^{\rm YM} \!\!&=&\!\!
    \frac{\tilde{P}^a_i \tilde{P}_a^i + B^a_i B_a^i}{2 \sqrt{\det q}}
    \,,\\
    \label{eq:Diffeomorphism constraint YM contribution - YM - classical}
    H^{\rm YM}_a \!\!&=&\!\!
    P^b_i F_{a b}^i
    \,,\\
    \label{eq:Gauss constraint - YM}
    G_i \!\!&=&\!\! - \partial_a P^a_i - g \tensor{f}{_i_j^k} A^j_a P^a_k
    \,,
\end{eqnarray}
with $\tilde{P}^a_i=P^a_i+\theta B^a_i$, and coupling constant $g$.
The gauge content of Yang--Mills theories on a fixed spacetime background is contained in the Gauss constraint $G_i$ whose algebra under the Poisson brackets is first-class,
\begin{equation}\label{eq:su(n) Lie}
    \{G_i[A_t^i],G_j[{\cal A}^j]\} = - G_i \left[g \tensor{f}{^i_j_k} A_t^j {\cal A}^k\right]\,,
\end{equation}
with structure \emph{constants} $\tensor{f}{^i_j_k}$, reflecting the underlying $\mathfrak{su}(n)$ Lie algebra and generating the corresponding gauge transformations: $\{A^i_a,G_i[{\cal A}^i]\} = \partial_a {\cal A}^i + g \tensor{f}{^i_j_k} A^j_a {\cal A}^k= \delta^{{\rm SU}(n)}_{{\cal A}} A_t^i$, $\{E^a_i,G_k[{\cal A}^k]\}= g \tensor{f}{^i^j_k} E^a_j {\cal A}^k= \delta^{{\rm SU}(n)}_{{\cal A}} E^a_i$.

The full Einstein--Yang--Mills system with a dynamical gravitational field results in a constraint algebra with additional Yang--Mills structure functions given by the components of the strength tensor:
\begin{subequations}\label{eq:Constraint algebra - EYM}
\begin{eqnarray}
    \{H_a[N^a],H_b[\epsilon^b]\}
    \!\!&=&\!\! - H_a\left[\mathcal{L}_{\vec{\epsilon}}N^a\right]
    - G_i\left[\epsilon^aN^b F^i_{ab}\right]\,\,\,\,
    \\
    \{H[N],H_a[\epsilon^a]\} \!\!&=&\!\! - H[\epsilon^a\partial_a N]
    + G_i\left[N\epsilon^aF^i_{0a}\right]
    \\
    \{H[N],H[\epsilon^0]\} \!\!&=&\!\! - H_a\left[q^{ab} \left(\epsilon^0\partial_bN-N\partial_b\epsilon^0\right)\right]\label{eq:Constraint algebra - EYM - HH}
    \\
    \{G_i[A_t^i],G_j[{\cal A}^j]\}
    \!\!&=&\!\! - G_i \left[g \tensor{f}{^i_j_k} A_t^j {\cal A}^k\right]\\
    \{H_a[N^a],G_i[{\cal A}^i]\} \!\!&=&\!\! 0
    \quad,\quad
    \{H[N],G_i[{\cal A}^i]\} = 0\,.
\end{eqnarray}
\end{subequations}
The algebra (\ref{eq:Constraint algebra - EYM}) has the same geometric interpretation of the hypersurface deformation algebra with additional deformations of the Yang--Mills-potential, and it describes the theory's covariance under SU($n$) transformations and spacetime diffeomorphisms.

It is important to note that this algebra is universal \cite{f(R)}: Regardless of the nature of new degrees of freedom contributing to the dynamics, whether they come from exotic matter or higher-curvature terms, the resulting constraints always satisfy the same algebra (\ref{eq:Constraint algebra - EYM}) ---with additional Gauss constraints if the new degrees of freedom correspond to new Yang--Mills fields.
New degrees of freedom imply only generic contributions to the vector and Gauss constraints respectively based on the rank of the new degrees of freedom as spatial and $\mathfrak{su}(n)$-valued tensors.
On the other hand, contributions to the Hamiltonian constraint are usually nontrivial; therefore, it is the Hamiltonian constraint that primarily governs the dynamics.

A consequence of the decomposition of spacetime tensors into spatial ones in the canonical formulation is that general covariance is not manifest as it is in the Lagrangian formulation: While the canonical gauge transformations generated by the vector and Gauss constraints are easily identified as spatial diffeomorphisms and SU($n$) transformations, those generated by the Hamiltonian constraint are not trivially evident.
Contrary to common belief, the hypersurface deformation algebra (\ref{eq:Hypersurface deformation algebra}) does not imply general covariance; it needs to be checked as an extra step \cite{EMGCov}.
In particular, we say that the spacetime is covariant if the canonical gauge transformation is equivalent to a diffeomorphism of the spacetime metric
\begin{equation}
    \delta_{\epsilon,{\cal A}} g_{\mu \nu} \big|_{\rm O.S.} =
    \mathcal{L}_{\xi} g_{\mu \nu} \big|_{\rm O.S.}
    \,,
    \label{eq:Spacetime covariance condition}
\end{equation}
where $\delta_{\epsilon,{\cal A}}\cdot=\{\cdot,H[\epsilon^0]+H_a[\epsilon^a]+G_i[{\cal A}^i]\}$ and "O.S." indicates an evaluation on shell, involving not only the vanishing of the constraints but also use of Hamilton's equations of motion in the right-hand-side for the time-derivatives, $\dot{q}_{ab}=\{q_{ab},H[N]+H_c[N^c]+G_i[A_t^i]\}$.
The gauge parameters $(\epsilon^0, \epsilon^a)$ in (\ref{eq:Spacetime covariance condition}) are related to the four-vector field $\xi^\mu$ that acts as the diffeomorphism generator by a change of basis from the coordinate frame to the Eulerian frame associated to the foliation:
\begin{eqnarray}
    \xi^\mu \!\!&=&\!\! \xi^t t^\mu + \xi^a s^\mu_a = \epsilon^0 n^\mu + \epsilon^a s^\mu_a
    \,,\nonumber
    \\
    \xi^t \!\!&=&\!\! \frac{\epsilon^0}{N}
    \quad,\quad
    \xi^a = \epsilon^a - \frac{\epsilon^0}{N} N^a
    \,.
\label{eq:Diffeomorphism generator projection}
\end{eqnarray}

The components of the spacetime metric in (\ref{eq:Spacetime covariance condition}) are given by those of the ADM line element (\ref{eq:ADM line element}), which are expressed in terms of $N$, $N^a$, and $q_{ab}$.
Therefore, the left-hand-side of (\ref{eq:Spacetime covariance condition}) requires gauge transformations of $N$ and $N^a$ which are not directly provided by the Poisson brackets because they do not have momenta.
Instead, the gauge transformations of the lapse and shift are derived from the condition that the equations of motion are gauge covariant, yielding \cite{pons1997gauge,salisbury1983realization,bojowald2018effective}
\begin{eqnarray}
    \delta_{\epsilon,{\cal A}}  N \!\!&=&\!\! \dot{\epsilon}^0 + \epsilon^a \partial_a N - N^a \partial_a \epsilon^0
    \,,
    \label{eq:Off-shell gauge transformations for lapse - intro}
    \\
    \delta_{\epsilon,{\cal A}} N^a \!\!&=&\!\! \dot{\epsilon}^a + \mathcal{L}_{\vec \epsilon} N^a
    + q^{a b} \left(\epsilon^0 \partial_b N - N \partial_b \epsilon^0 \right)
    \,.
    \label{eq:Off-shell gauge transformations for shift - intro}
\end{eqnarray}
Analogously, the Yang--Mills potential transforms as
\begin{equation}
    \delta_{\epsilon,{\cal A}} A_t^i =
    \dot{\cal A}^i
    + g \tensor{f}{^i_j_k} A_t^j {\cal A}^k
    + F^i_{ab} \epsilon^a N^b
    + F^i_{0a} \left(\epsilon^0 N^a - N \epsilon^a\right)
    \,.\label{eq:A_t gauge transf - EYM}
\end{equation}

In other words, the canonical gauge transformations are given by the combined operation
\begin{eqnarray}
    {\cal O} \!\!&\to&\!\! {\cal O} + \{{\cal O},H[\epsilon^0]+H_a[\epsilon^a]+G_i[{\cal A}^i]\}
    \,,\nonumber\\
    N \!\!&\to&\!\! N + \delta_{\epsilon,{\cal A}} N
    \quad,\quad
    N^a \to N^a + \delta_{\epsilon,{\cal A}} N^a
    \,,\nonumber\\
    A_t^i \!\!&\to&\!\! A_t^i + \delta_{\epsilon,{\cal A}} A_t^i
    \,,
\end{eqnarray}
for all phase-space functions ${\cal O}$.
This allows for the evaluation of the covariance equation (\ref{eq:Spacetime covariance condition}), whose $tt$ and $ta$ components are automatically satisfied by the gauge transformations of the lapse and shift, (\ref{eq:Off-shell gauge transformations for lapse - intro}) and (\ref{eq:Off-shell gauge transformations for shift - intro}), provided the covariance equation of the spatial components, $\delta_{\epsilon,{\cal A}} q_{a b} |_{\rm O.S.} = \mathcal{L}_{\xi} q_{a b} |_{\rm O.S.}$, is satisfied too \cite{EMGCov}.
The constraints of general relativity, even in the presence of matter fields, indeed satisfy this equation; therefore, it can be seen as an identity of the theory.

Eq.~(\ref{eq:A_t gauge transf - EYM}) can be rewritten as a linear combination of a spacetime diffeomorphism and a SU($n$) transformation of the time-component of the connection,
\begin{equation}
    \delta_{\epsilon,{\cal A}} A_t^i =
    \mathcal{L}_\xi A_t^i
    + \delta^{{\rm SU}(n)}_{{\cal A} - \xi^\sigma A_\sigma} A_t^i
    \,,
\end{equation}
by use of $F_{ta}^i=N F^i_{0a} + N^bF^i_{ba}$ and $F_{ta}^i=\dot{A}_a^i-\partial_aA_t^i + g \tensor{f}{^i_j_k} A^j_t A^k_a$.
It follows that the canonical gauge transformations of the strength tensor field and the connection must be given by
\begin{eqnarray}\label{eq:Strength tensor covariance condition}
    \delta_{\epsilon,{\cal A}} F_{\mu\nu}^i\big|_{\rm O.S.} \!\!&=&\!\!
    \mathcal{L}_\xi F_{\mu\nu}
    + \delta^{{\rm SU}(n)}_{{\cal A} - \xi^\sigma A_\sigma} F_{\mu\nu}^i\big|_{\rm O.S.}
    \,,\\
    \delta_{\epsilon,{\cal A}} A_\mu^i\big|_{\rm O.S.} \!\!&=&\!\!
    \mathcal{L}_\xi A_\mu^i
    + \delta^{{\rm SU}(n)}_{{\cal A} - \xi^\sigma A_\sigma} A_\mu^i\big|_{\rm O.S.}\,.\label{eq:YM 1-form covariance condition}
\end{eqnarray}
These covariance equations are also satisfied by the classical constraints and are therefore identities of the theory too.

Finally, we point out that the Einstein--Yang--Mills system contains the following (nonlocal) Dirac observables
\begin{equation}\label{eq:YM observable}
    \mathfrak{G}_i[\Omega^i] = - \int{\rm d}^3x\;\Omega^i \tensor{f}{_i_j^k} A^j_a P^a_k\,,
\end{equation}
where $\Omega^i$ is a $\mathfrak{su}(n)$-valued constant.
As a local function, $\mathfrak{G}^{\rm YM}_i(x)$ has the following brackets with the constraints,
\begin{eqnarray}
    \{\mathfrak{G}_i,H[N]\} \!\!&=&\!\!
    \partial_a \left(\epsilon^{a b c} \tensor{f}{_i_j^k} A^j_b B_c^k N\right)
    \,,\\
    \{\mathfrak{G}_i,H_a[N^a]\} \!\!&=&\!\!
    \partial_a\left(\mathfrak{G}_i N^a + \tensor{f}{_i_j^k} A^j_c P^a_k N^c\right)
    \,,\\
    \{\mathfrak{G}_i,G_i[A_t^i]\} \!\!&=&\!\!
    - \partial_a \left(\tensor{f}{_i_j^k} A_t^j P^a_k\right)
    - \tensor{f}{_i_n^k} A_t^n G_k
    \,;\quad
\end{eqnarray}
therefore, the functional (\ref{eq:YM observable}) commutes with the constraints on shell up to boundary terms.
Furthermore, such boundary terms define a conserved densitized four-current with components
\begin{eqnarray}
    J_i^t\!\!&=& - \mathfrak{G}_i
    \!\!
    \,,\\
    J_i^a\!\!&=&\!\! - \epsilon^{a b c} \tensor{f}{_i_j^k} A^j_b B_c^k N
    - \mathfrak{G}_i N^a
    + \tensor{f}{_i_j^k} \left(A_t^j-N^c A^j_c\right) P^a_k
    ,\nonumber
\end{eqnarray}
because the brackets above imply that, on shell, $\dot{J}_i^t=-\partial_a J_i^a$ and hence $\partial_\mu J^\mu=\nabla_\mu J^\mu_i=0$.
Finally, the bracket of the observable's local function with itself forms the $\mathfrak{su}(n)$ algebra:
\begin{equation}
    \{\mathfrak{G}_i(x),\mathfrak{G}_j(y)\} = \tensor{f}{_i_j^k} \mathfrak{G}_k \delta^3(x-y)
    \,.
\end{equation}
For the special case of U(1), the observable is instead $\mathfrak{G}_{\rm U(1)}^a[\Omega_a] = \int{\rm d}^3x\Omega_a P^a$ with constant $\Omega_a$.

The physical fields of the Einstein--Yang--Mills system are given by the metric, the strength tensor, and the connection because these are the ones that are directly probed by test particles:
Consider a point particle with mass $m$ and SU($n$) charge $\alpha^i$ (valued in $\mathfrak{su}(n)$) in the presence of an external Yang--Mills field in a curved spacetime.
The particle's four-velocity $u^\mu$ and its charge $\alpha^i$ evolve in time according to the Wong equations
\begin{equation}\label{eq:YM Lorentz force}
    m u^\nu \nabla_\nu u^\mu = g \alpha^i F_i^{\mu\nu} u_\nu
\end{equation}
and
\begin{equation}\label{eq:Wong charge evolution}
    \frac{{\rm d} \alpha^i}{{\rm d \tau}} = - g \tensor{f}{^i_j_k} A^j_\mu u^\mu \alpha^k
    \,.
\end{equation}

The Wong equations are SU($n$)-covariant if the charge SU($n$)-transforms as
\begin{equation}\label{eq:Test particle SU(n)-gauge transf}
    \alpha^i \to \alpha^i+g\tensor{f}{^i_j_k} \alpha^j \omega^k\,,
\end{equation}
and the four-velocity remains invariant.
On the other hand, the Wong equations are covariant under coordinate transformations if the four-velocity transforms as a spacetime vector field and the charge is an invariant function of proper time.
This defines the transformation properties of the particle variables $u^\mu$ and $\alpha^i$.

Finally, the uniqueness theorem of canonical general relativity in vacuum \cite{hojman1976geometrodynamics,kuchar1974geometrodynamics} states that if the constraint $H$ of a gravitational system with the canonical pair $(q_{ab},p^{cd})$ is required to satisfy, together with the vector constraint, the hypersurface deformation algebra (\ref{eq:Hypersurface deformation algebra}), specifically with the structure function $q^{ab}$ being the inverse of the configuration variable $q_{ab}$ ---or related by a canonical transformation ---, and assuming $H$ depends on up to second-order spatial derivatives of $q_{ab}$ as well as assuming time-reversal and parity invariance, then the Hamiltonian constraint $H$ is given uniquely by that of general relativity up to the choice of Newton's and the cosmological constants. This result rules out the possibility of modified gravity to this derivative order.
However, this (often implicit) assumption that the spatial metric is precisely the canonical configuration variable $q_{ab}$ (up to canonical transformations) can be dropped so that the uniqueness theorems do not necessarily apply, although this leaves the relation between the phase-space variables to the spacetime metric undefined.
Emergent modified gravity \cite{EMG,EMGCov} provides a process to regain the spacetime geometry using the concept of structure functions in the constraint algebra.
Extending these arguments to the Yang--Mills fields and the new structure functions results in the more encompassing emergent field theory, which we develop in the next section.

\section{Emergent fields}
\label{sec:Emergent fields}

The first step in every Hamiltonian system is to define its phase-space structure.
We define the gravitational and Yang--Mills fields as the canonical pairs $(q_{ab},p^{cd})$ and $(A^i_a,P^b_j)$, respectively, satisfying the basic Poisson brackets.
The dynamics are then generated by the set of first-class constraints $\tilde{H}$, $H_a$, and $G_i$ as functions of the phase space.
Having retained the classical form of $H_a$ and $G_i$, the SU($n$) and spatial diffeomorphism transformations of the phase-space variables are retained; in particular, their density weights permit invariant spatial integrations for suitable functions of the phase space without relying on a spatial metric.
However, the Hamiltonian constraint $\tilde{H}$ is not necessarily equal to its classical expression, but is required to satisfy the Einstein--Yang--Mills algebra
\begin{subequations}\label{eq:Constraint algebra - EMGFT}
\begin{eqnarray}
    \{H_a[N^a],H_b[\epsilon^b]\}
    \!\!&=&\!\! - H_a\left[\mathcal{L}_{\vec{\epsilon}}N^a\right]
    - G_i\left[\epsilon^aN^b F^i_{ab}\right]
    \quad\\
    \{\tilde{H}[N],H_a[\epsilon^a]\} \!\!&=&\!\! - \tilde{H}[\epsilon^a\partial_aN]
    + G_i\left[N\epsilon^a\tilde{F}^i_{0a}\right]
    \label{eq:H,Ha bracket - EMGFT}\\
    \{\tilde{H}[N],\tilde{H}[\epsilon^0]\} \!\!&=&\!\! - H_a\left[\tilde{q}^{ab} \left(\epsilon^0\partial_bN-N\partial_b\epsilon^0\right)\right]
    \\
    \{G_i[A_t^i],G_j[{\cal A}^j]\}
    \!\!&=&\!\! - G_i \left[g \tensor{f}{^i_j_k} A_t^j {\cal A}^k\right]\\
    \{H_a[N^a],G_i[{\cal A}^i]\} \!\!&=&\!\! 0
    \quad,\quad
    \{\tilde{H}[N],G_i[{\cal A}^i]\} = 0\,,
\end{eqnarray}
\end{subequations}
up to the structure functions $\tilde{q}^{ab}$ and $\tilde{F}^i_{0a}$ which are not necessarily equal to their classical expressions.
Furthermore, we require that the structure functions are not related to their classical expressions by canonical transformations that leave the vector and Gauss constraints invariant; if they were related by a canonical transformation, they would describe the same physical system but in different phase-space coordinates despite featuring different forms of the Hamiltonian constraint and the structure functions, and hence they would be equivalent to the classical theory ---notice that all of the covariance conditions detailed below are formulated in terms of Poisson brackets, whose results are invariant to canonical transformations.
We also refer to this requirement as the anomaly-freedom condition.

This modified constraint algebra defines the gauge transformations of the lapse and shift which are now respectively given by (\ref{eq:Off-shell gauge transformations for lapse - intro}) and (\ref{eq:Off-shell gauge transformations for shift - intro}) with $q^{ab}$ replaced by $\tilde{q}^{ab}$.
Similarly, the gauge transformation of the Yang--Mills potential is given by (\ref{eq:A_t gauge transf - EYM}) with $F_{0a}^i$ replaced by $\tilde{F}_{0a}^i$.

While the covariance equations (\ref{eq:Spacetime covariance condition}), (\ref{eq:Strength tensor covariance condition}), and (\ref{eq:YM 1-form covariance condition}) are identities of the classical system, for modified theories they become conditions because the preservation of the algebra (\ref{eq:Constraint algebra - EMGFT}) does not guarantee that the modified constraints generate gauge transformations compatible with the covariance equations \cite{EMGCov,EMGPF,EMGscalar,EMGEM}.
In particular, compatibility of the modified gauge transformations of the Lagrange multipliers with the covariance conditions imply that the inverse of the structure function $\tilde{q}_{ab}$ must be used as the spatial metric in the line element (\ref{eq:ADM line element}) instead of the configuration variable $q_{ab}$.
Similarly, the structure function $\tilde{F}_{0a}^i$ must be used as the electric components the strength tensor field.
This procedure results in the emergence of the spacetime metric $\tilde{g}_{\mu\nu}$ and the strength tensor field $\tilde{F}_{\mu\nu}^i$ ---which implies an emergent electric field $\tilde{E}^a_i$ ---as functions of the phase space from the gauge symmetry described by the constraint algebra and its consistency with covariance.
The use of the emergent structure functions and the modified constraints in place of the corresponding classical expressions in the covariance conditions (\ref{eq:Spacetime covariance condition}), (\ref{eq:Strength tensor covariance condition}), and (\ref{eq:YM 1-form covariance condition}) places restrictions on their phase-space dependence beyond what anomaly-freedom by itself implies as we detail below.
Electromagnetism, as an Abelian subsystem, has no structure constants and hence the charge-evolution equation (\ref{eq:Wong charge evolution}) is trivial; therefore, the electromagnetic 1-form $A_\mu$ does not enter direct observations because it does not interact with test particles.
Consequently, electromagnetic covariance does not require imposing the condition (\ref{eq:YM 1-form covariance condition}) and hence is less restricted compared to its SU($n>1$) cousins.

Because no modifications are introduced to the vector and Gauss constraints, only the normal transformations in the covariance conditions will differ from their classical versions.
Therefore, the spacetime covariance condition reduces to \cite{EMGCov}
\begin{equation}\label{eq:Spacetime cov condition - reduced - EMGFT}
    \frac{\partial\{\tilde{q}^{ab},\tilde{H}[\epsilon^0]\}}{\partial (\partial_{c_1}\epsilon^0)}\bigg|_{\rm O.S.}
    = \frac{\partial\{\tilde{q}^{ab},\tilde{H}[\epsilon^0]\}}{\partial (\partial_{c_1}\partial_{c_2}\epsilon^0)}\bigg|_{\rm O.S.}
    = \dotsi = 0\,,
\end{equation}
and the Yang--Mills covariance conditions to
\begin{equation}\label{eq:Vector potential cov conditions - reduced - EMGFT}
    \frac{\partial\tilde{H}}{\partial (\partial_{c_1}P_a^i)}\bigg|_{\rm O.S.}
    =  \frac{\partial\tilde{H}}{\partial (\partial_{c_1}\partial_{c_2}P_a^i)}\bigg|_{\rm O.S.}
    = \dotsi = 0
\end{equation}
for the vector potential,
\begin{eqnarray}\label{eq:Magnetic cov conditions - reduced - EMGFT - 1}
    \tilde{F}_{0 a}^i \big|_{\rm O.S.} \!\!&=&\!\! \{A_{a}^i,\tilde{H}[1]\}
    \big|_{\rm O.S.}
\end{eqnarray}
for the magnetic components of the strength tensor, and
\begin{eqnarray}\label{eq:Electric cov conditions - reduced - EMGFT}
    \frac{\partial\{\tilde{F}_{0 a}^i,\tilde{H}[\epsilon^0]\}}{\partial (\partial_c \epsilon^0)}
    \!\!&=&\!\! \tilde{q}^{cb} F_{b a}^i
    \bigg|_{\rm O.S.}
    \,,\\
    \frac{\partial\{\tilde{F}^i_{0a},\tilde{H}[\epsilon^0]\}}{\partial (\partial_{c_1}\partial_{c_2}\epsilon^0)}\bigg|_{\rm O.S.} \!\!&=&\!\! \frac{\partial\{\tilde{F}^i_{0a},\tilde{H}[\epsilon^0]\}}{\partial (\partial_{c_1}\partial_{c_2}\partial_{c_3}\epsilon^0)}\bigg|_{\rm O.S.} = \dotsi = 0\,,\qquad
\end{eqnarray}
for the electric components.
Eq.~(\ref{eq:Vector potential cov conditions - reduced - EMGFT}) is also implied by the covariance condition on $F_{ab}^i$ even in the U(1) case; therefore, the U(1) connection $A_\mu$ too has a covariant transformation of the form (\ref{eq:YM 1-form covariance condition}), but this arises as a consequence of requiring the magnetic field be covariant and not from a direct condition on $A_\mu$.

Noting that the phase-space functional
\begin{equation}
    D_a[\epsilon^a] = H_a[\epsilon^a]+G_i[\epsilon^a A_a^i]
\end{equation}
generates spatial diffeomorphisms of all phase-space variables, it follows from the bracket (\ref{eq:H,Ha bracket - EMGFT}) that the electric structure functions are given by
\begin{equation}
    \tilde{F}_{0a}^i = \frac{1}{\epsilon^0} \{A_a^i,\tilde{H}[\epsilon^0]\}
\end{equation}
off shell, which is in agreement with the on-shell covariance condition (\ref{eq:Magnetic cov conditions - reduced - EMGFT - 1}) necessary for $\tilde{F}^i_{ta}=N\tilde{F}^i_{0a}+N^b\tilde{F}^i_{ba}$ be related to the vector potential by a time derivative.
The gravitational structure function $\tilde{q}^{ab}$, on the other hand, has no shortcut derivation and the full algebra must be explicitly imposed to obtain it as we show in the next section for spherically symmetric systems.

The emergent fields, rather than the direct phase-space variables, must be considered as the physical manifestations of gravity and Yang--Mills forces because their consistent covariant transformations allow us to identify them as spacetime tensor fields and can be consistently used in the Wong equation (\ref{eq:YM Lorentz force}) which dictates the dynamics of test particles.
This leads to an explicit distinction between the fundamental gravitational field, described by the canonical pair $(q_{ab},p^{cd})$, and the spacetime, described by the emergent metric field $\tilde{g}_{\mu\nu}$, as well as between the fundamental Yang--Mills field, described by the canonical pair $(A^i_a,P^b_j)$, and the Yang--Mills force, described by the emergent strength tensor field $\tilde{F}_{\mu\nu}^i$.

As a final point, if one considers additional matter fields into the theory that present manifestations independently from the spacetime metric or the Yang--Mills forces, then one has to make sure that such manifestations of the matter fields are covariant too.
For instance, if the matter field in consideration is physically described by some $\mathfrak{su}(n)$-valued spacetime tensor field $f$, then one has to apply the matter covariance condition on this field too:
\begin{equation}\label{eq:Covariance condition - general}
    \delta_{\epsilon,{\cal A}} f |_{\rm O.S.} = \mathcal{L}_\xi f + \delta^{{\rm SU}(n)}_{{\cal A} - \xi^\sigma A_\sigma} f |_{\rm O.S.}
    \,.
\end{equation}
Scalar and fermionic matter, as well as perfect fluids, contribute to the Hamiltonian and vector constraints, and, if they are SU($n$)-charged, to the Gauss constraint, but their presence does not change the Einstein--Yang--Mills constraint algebra or introduce additional constraints.
$\mathfrak{su}(n)$-valued scalar matter was studied in detail in \cite{EMGscalar}, where its configuration variable $\phi_i$ replaces $f$ in the covariance condition (\ref{eq:Covariance condition - general}), which reduces to
\begin{equation}\label{eq:Vector potential cov conditions - EMGFT}
    \frac{\partial\tilde{H}}{\partial (\partial_{c_1}P_\phi^i)}\bigg|_{\rm O.S.}
    =  \frac{\partial\tilde{H}}{\partial (\partial_{c_1}\partial_{c_2}P_\phi^i)}\bigg|_{\rm O.S.}
    = \dotsi = 0\,,
\end{equation}
where $P_\phi^i$ is the corresponding conjugate momentum.
Neutral perfect fluid has been studied in detail as well in \cite{EMGPF}, where its co-velocity $u_\mu$ replaces $f$ in the covariance condition (\ref{eq:Covariance condition - general}); however, $u_\mu$ is a function of the fluid's configuration variables and momenta, resulting in lengthier expressions for the covariance conditions.

\section{Spherical symmetry}
\label{sec:Spherically symmetric system}

\subsection{Kinematics}

\subsubsection{Phase-space structure}

In the spherically symmetric theory, the spacetime metric takes the general form
\begin{equation}
    {\rm d} s^2 = - N^2 {\rm d} t^2 + \tilde{q}_{x x} ( {\rm d} x + N^x {\rm d} t )^2 + \tilde{q}_{\vartheta \vartheta} {\rm d} \Omega^2
    \label{eq:ADM line element - spherical}
    \,.
\end{equation}
The classical spatial metric components can be written in terms of the classical radial and angular densitized triads ${\cal E}^x$ and ${\cal E}^\varphi$, respectively, $q_{xx} = ({\cal E}^\varphi)^2/{\cal E}^x$ and $q_{\vartheta\vartheta} = {\cal E}^x$, but $\tilde{q}_{xx}$ and $\tilde{q}_{\vartheta \vartheta}$ may acquire more general forms in the modified theory.

In spherical symmetry, only the radial component of the electromagnetic vector potential $A_x$ and of its conjugate momentum ${\cal P}^x$ are nontrivial.
Therefore, the only nontrivial component of the classical strength tensor is given by
\begin{equation}
    F_{0x}=\frac{{\cal E}^\varphi}{({\cal E}^x)^{3/2}} \frac{{\cal P}^x}{4\pi} =: \frac{\bar{F}_{0x}}{4\pi}\,,
\end{equation}
where the $4\pi$ rescaling of $\bar{F}_{0x}$, and of other quantities in what follows, turns out to be computationally convenient to absorb similar terms that result from angular integrations.
On the other hand, a spherically symmetric SU(2) vector potential $W^i_a$ has the richer form \cite{bojowald2000symmetry}
\begin{eqnarray}
    W_a^i {\rm d}x^a\tau_i \!\!&=&\!\! \frac{W_x(x)}{\sqrt{4\pi}} \tau_3 {\rm d} x
    + \frac{W_1(x)\tau_1+W_2(x)\tau_2}{\sqrt{8\pi}} {\rm d} \vartheta
    \nonumber\\
    \!\!&&\!\!
    + \frac{W_1(x) \tau_2 - W_2(x) \tau_1}{\sqrt{8\pi}} \sin \vartheta {\rm d} \varphi
    \nonumber\\
    \!\!&&\!\!
    + \tau_3 \cos \vartheta {\rm d} \varphi
    \,,\label{eq:SU(2) connection - spherical}
\end{eqnarray}
with momenta $P^a_i$ given by
\begin{eqnarray}
    P^a_i \tau^i \partial_a \!\!&=&\!\! \frac{P^x(x)}{\sqrt{4\pi}} \tau_3 \sin\vartheta \partial_x
    + \frac{P_1(x)\tau_1+P_2(x)\tau_2}{\sqrt{8\pi}} \sin\vartheta \partial_\vartheta
    \,\nonumber\\
    \!\!&&\!\!
    + \frac{P_1(x) \tau_2 - P_2(x) \tau_1}{\sqrt{8\pi}} \partial_\varphi
    \,,
\end{eqnarray}
where $\tau_i$ are $\mathfrak{su}(2)$ the generators.
(The $W_\varphi^3$ term in (\ref{eq:SU(2) connection - spherical}) implies a magnetic-monopole term in the strength tensor (\ref{eq:W - magnetic components - spherical}) below without Dirac strings. This can result, for instance, if the SU(2) Yang--Mills field couples a Higgs field which spontaneously breaks it down to an effective U(1) group via a Higgs-like mechanism \cite{tHooft,Polyakov}.)
We correspondingly rescale the only nontrivial SU(2) potential as $W^3_t=W_t/\sqrt{4\pi}$.
The nontrivial spatial components of the corresponding strength tensor are given by
\begin{eqnarray}\label{eq:W - magnetic components - spherical}
    {\cal W}^1_{x\vartheta} \!\!&=&\!\! \frac{W_1' - {\not\!g} W_x W_2}{\sqrt{8\pi}}
    =: \frac{{\not\!B}^\varphi_1}{\sqrt{8\pi}}
    \,,\\
    {\cal W}^2_{x\vartheta} \!\!&=&\!\! \frac{W_2' + {\not\!g} W_x W_1}{\sqrt{8\pi}}
    =: \frac{{\not\!B}^\varphi_2}{\sqrt{8\pi}}
    \,,\nonumber\\
    {\cal W}^1_{x\varphi} \!\!&=&\!\! - \sin\vartheta {\cal W}^2_{x\vartheta}
    \qquad,\quad
    {\cal W}^2_{x\varphi} = \sin\vartheta {\cal W}^1_{x\vartheta}
    \,,\nonumber\\
    {\cal W}^3_{\vartheta\varphi} \!\!&=&\!\! \frac{\sin\vartheta}{\sqrt{4\pi}} \left(\frac{{\not\!g}}{2} \left((W_1)^2 + (W_2)^2\right)-\sqrt{4\pi}\right)
    =: \frac{\sin\vartheta}{\sqrt{4\pi}} {\not\!B}^x
    \,,\nonumber
\end{eqnarray}
where ${\not\!g}=g/\sqrt{4\pi}$ and the prime denotes a radial derivative.
The non-vanishing components of the magnetic field are therefore given by
\begin{equation}
    B^x_3 = {\cal W}_{\vartheta \varphi}^3
    \quad,\quad
    B^\vartheta_i = - {\cal W}_{x \varphi}^i
    \quad,\quad
    B^\varphi_i = {\cal W}_{x \vartheta}^i
    \,,
\end{equation}
and the nontrivial normal components of the strength tensor by
\begin{eqnarray}
    {\cal W}^3_{0x} \!\!&=&\!\! \frac{{\cal E}^\varphi}{({\cal E}^x)^{3/2}} \frac{P^x+\theta {\not\!B}^x}{\sqrt{4\pi}}
    =: \frac{\bar{{\cal W}}^3_{0x}}{\sqrt{4\pi}}
    \,,\\
    {\cal W}^1_{0\vartheta} \!\!&=&\!\! \frac{\sqrt{{\cal E}^x}}{{\cal E}^\varphi} \frac{P_1+\theta {\not\!B}^\varphi_2}{\sqrt{8\pi}}
    =: \frac{\bar{{\cal W}}^1_{0\vartheta}}{\sqrt{8\pi}}
    \,,\nonumber\\
    {\cal W}^2_{0\vartheta} \!\!&=&\!\! \frac{\sqrt{{\cal E}^x}}{{\cal E}^\varphi} \frac{P_1-\theta {\not\!B}^\varphi_1}{\sqrt{8\pi}}
    =: \frac{\bar{{\cal W}}^2_{0\vartheta}}{\sqrt{8\pi}}
    \,,\nonumber\\
    {\cal W}^1_{0\varphi} \!\!&=&\!\! - \sin\vartheta {\cal W}^2_{0\vartheta}
    \qquad,\quad
    {\cal W}^2_{0\varphi} = \sin\vartheta {\cal W}^1_{0\vartheta}
    \,.\nonumber
\end{eqnarray}

Finally, the perfect fluid is described by the co-velocity
\begin{equation}
    u_\mu{\rm d}x = - \frac{1}{\sqrt{4\pi}} \left(\partial_\mu T + \frac{P_X}{P_T} \partial_\mu X\right)
\end{equation}
with two configuration variables $T$ and $X$ and their respective momenta $P_T$ and $P_X$.
And scalar matter is easily described by the configuration variable $\phi(x)$ and its momentum $P_\phi(x)$ appropriately rescaled by factors of $4\pi$.

Based on the classical system, we define the phase-space structure as follows.
The gravitational field is given by the canonical pairs $(K_x,{\cal E}^x)$ and $(K_\varphi,{\cal E}^\varphi)$, where $K_x$ and $K_\varphi$ are related to the radial and angular components of the extrinsic curvature in the classical theory \cite{bojowald2000symmetry,bojowald2004spherically}; the electromagnetic field is given by the pair $(A_x,{\cal P}^x)$; the SU(2) Yang--Mills field by the pairs $(W_x,P^x)$, $(W_1,P_1)$, and $(W_2,P_2)$; the scalar matter field by the pair $(\phi,P_\phi)$; and the perfect fluid by the pairs $(T,P_T)$ and $(X,P_X)$.
Each canonical pair $(Q,P)$ satisfies the basic Poisson bracket $\{Q(x),P(y)\}=\delta(x-y)$.

\subsubsection{Covariance conditions}
\label{sec:Covariance conditions - spherical}

In spherical symmetry, the vector constraint reduces to
\begin{eqnarray}
    H_x \!\!&=&\!\!
    {\cal E}^\varphi K_\varphi'
    - K_x ({\cal E}^x)'
    + P_\phi \phi'
    + P_X X' + P_T T'
    \nonumber\\
    \!\!&&\!\!
    + P_1W_1'+P_2W_2'
    + {\not\!g} W_x (P_2W_1-P_1W_2)
    \,,
    \label{eq:Diffeomorphism constraint - Gravity - spherical}
\end{eqnarray}
which is independent of electromagnetic variables because its magnetic field vanishes.
The U(1) Gauss constraint reduces to
\begin{equation}\label{eq:Gauss constraint - spherical}
    G^{\rm U(1)} = - ({\cal P}^x)'\,,
\end{equation}
and the SU(2) Gauss constraint to
\begin{eqnarray}
    G^{\rm SU(2)} = - (P^x)'
    - {\not\!g} \left(P_2 A_1
    - P_1 A_2\right)\,,
\end{eqnarray}
which now behaves as an effective U(1) generator around the third axis of the internal SU(2) space.
(Charged matter contributes to the Gauss constraints, but we consider only neutral matter in the following.)
The Dirac observable (\ref{eq:YM observable}) reduces to
\begin{equation}\label{eq:YM observable - spherical}
    \mathfrak{G}[\Omega] = - \int{\rm d}x\;\Omega \left(P_2 W_1
    - P_1 W_2\right)\,,
\end{equation}
with arbitrary constant $\Omega$, and the U(1) observable to
\begin{equation}\label{eq:Local charge observable - U(1)}
    {\cal Q}_{\rm U(1)} = - {\cal P}^x\,,
\end{equation}
associated to the electric charge of the system.

In the modified theory, we retain the classical form of the vector constraint (\ref{eq:Diffeomorphism constraint - Gravity - spherical}) and of the Gauss constraint (\ref{eq:Gauss constraint - spherical}), but consider a modified Hamiltonian $\tilde{H}$ and impose the spherical Einstein--Yang--Mills algebra given by\footnote{This corrects a mistaken omission of the Gauss constraint contribution in (\ref{eq:Hypersurface deformation algebra - HHa - spherical - modified}) that was made in \cite{EMGEM}.}
\begin{subequations}\label{eq:Hypersurface deformation algebra - spherical - modified}
\begin{eqnarray}
    \{ H_x [ N^x] , H_x [ \epsilon^x ] \} \!\!&=&\!\! - H_x [\epsilon^x (N^x)' - N^x (\epsilon^x)']
    \,,\qquad
    \label{eq:Hypersurface deformation algebra - HaHa - spherical - modified}
    \\
    \{ \tilde{H} [ N ] , H_x [ \epsilon^x]\} \!\!&=&\!\! - \tilde{H} [ \epsilon^x N' ]
    + G^{\rm U(1)} \left[N\epsilon^x\tilde{\bar{F}}_{0x}\right]
    \nonumber\\
    \!\!&&\!\!
    + G^{\rm SU(2)} \left[N\epsilon^x\tilde{\bar{{\cal W}}}^3_{0x}\right]
    \,,\label{eq:Hypersurface deformation algebra - HHa - spherical - modified}\\
    \{ \tilde{H} [ N ] , \tilde{H} [ \epsilon^0 ] \} \!\!&=&\!\! - H_x [ \tilde{q}^{xx} ( \epsilon^0 N' - N (\epsilon^0)' )]
    \,,\qquad
    \label{eq:Hypersurface deformation algebra - HH - spherical - modified}
\end{eqnarray}
\end{subequations}
as well as the vanishing brackets of the Gauss constraints with all constraints, with structure functions $\tilde{q}^{xx}$, $\tilde{\bar{F}}_{0x}$, and $\tilde{\bar{{\cal W}}}^3_{0x}$ not necessarily equal to the classical $q^{xx}$, $\bar{F}_{0x}$, and $\bar{{\cal W}}^3_{0x}$.

To derive an explicit expression for the constraint and the structure function, we define an ansatz for $\tilde{H}$ as a function of the phase-space variables and their derivatives up to second order.
The condition of anomaly-freedom of the constraint algebra (\ref{eq:Hypersurface deformation algebra - spherical - modified}) places strong restrictions on this ansatz, and the bracket (\ref{eq:Hypersurface deformation algebra - HH - spherical - modified}) determines $\tilde{q}^{xx}$.
The inverse of the modified structure function $\tilde{q}_{xx}=1/\tilde{q}^{xx}$ is then used in the line element (\ref{eq:ADM line element - spherical}).
Unlike the radial component, the modified angular component $\tilde{q}^{\vartheta \vartheta}$ cannot be derived from the constraint algebra because the angular components of the vector constraint trivialize due to the underlying spherical symmetry.
The modified angular structure function must then be chosen using phenomenological considerations; here we will take it as an undetermined function of the radial triad $\tilde{q}^{\vartheta \vartheta}({\cal E}^x)$.
Similarly, the structure functions corresponding to the radial components of the electric fields $\bar{F}_{0x}$ and $\tilde{\bar{{\cal W}}}^3_{0x}$ are determined by the bracket (\ref{eq:Hypersurface deformation algebra - HHa - spherical - modified}) and given by
\begin{eqnarray}
    \tilde{\bar{F}}_{0 x} \!\!&=&\!\!
    \frac{\partial\tilde{H}}{\partial {\cal P}^x}
    \,,\label{eq:F0x - EMGFT}\\
    \tilde{\bar{{\cal W}}}_{0 x}^3 \!\!&=&\!\!
    \frac{\partial\tilde{H}}{\partial P^x}
    \,,\label{eq:W0x3 - EMGFT}
\end{eqnarray}
while the angular components $\tilde{\bar{{\cal W}}}^i_{0\vartheta}$ and $\tilde{\bar{{\cal W}}}^i_{0\varphi}$, are absent from the symmetry reduced algebra.
However, as we show below, these angular components can be derived from the covariance conditions.

Having identified the emergent fields, we can now impose the covariance conditions.
First, if the algebra (\ref{eq:Hypersurface deformation algebra - spherical - modified}) holds, then the canonical gauge transformation of the Lagrange multipliers become
\begin{eqnarray}
    \delta_{\epsilon,{\cal A},{\cal B}} N \!\!&=&\!\! \dot{\epsilon}^0 + \epsilon^x N' - N^x (\epsilon^0)'
    \,,
    \label{eq:Off-shell gauge transformations for lapse - EMG- spherical}
    \\
    \delta_{\epsilon,{\cal A},{\cal B}} N^x \!\!&=&\!\! \dot{\epsilon}^x + \epsilon^x (N^x)' - N^x (\epsilon^x)
    \nonumber\\
    \!\!&&\!\!
    + \tilde{q}^{xx} \left(\epsilon^0 N' - N (\epsilon^0)' \right)
    \,,\\
    \delta_{\epsilon,{\cal A},{\cal B}} A_t \!\!&=&\!\! \dot{\cal A} + \tilde{\bar{F}}_{0x}\left(\epsilon^0 N^x-N\epsilon^x\right)
    \,,\\
    \delta_{\epsilon,{\cal A},{\cal B}} W_t \!\!&=&\!\! \dot{\cal B} + \tilde{\bar{{\cal W}}}^3_{0x}\left(\epsilon^0 N^x-N\epsilon^x\right)
    \,,
    \label{eq:Off-shell gauge transformations for shift - EMG - spherical}
\end{eqnarray}
which are in agreement with the corresponding components of the covariance conditions (\ref{eq:Spacetime covariance condition}) and (\ref{eq:YM 1-form covariance condition}) upon symmetry reduction as well as using $\tilde{q}^{xx}$, $\tilde{\bar{F}}_{0x}$, and $\tilde{\bar{{\cal W}}}^3_{0x}$ instead of the classical expressions.
The covariance condition of the spacetime structure function (\ref{eq:Spacetime cov condition - reduced - EMGFT}) reduces to
\begin{eqnarray}\label{eq:Spacetime cov condition - reduced - EMGFT - spherical xx}
    \frac{\partial\{\tilde{q}^{xx},\tilde{H}[\epsilon^0]\}}{\partial (\epsilon^0)'}
    = \frac{\partial\{\tilde{q}^{xx},\tilde{H}[\epsilon^0]\}}{\partial (\epsilon^0)''}
    = \dotsi = 0\,,\\
    \label{eq:Spacetime cov condition - reduced - EMGFT - spherical thth}
    \frac{\partial\{\tilde{q}^{\vartheta\vartheta},\tilde{H}[\epsilon^0]\}}{\partial (\epsilon^0)'}
    = \frac{\partial\{\tilde{q}^{\vartheta\vartheta},\tilde{H}[\epsilon^0]\}}{\partial (\epsilon^0)''}
    = \dotsi = 0\,,
\end{eqnarray}
while the Yang--Mills covariance conditions (\ref{eq:Vector potential cov conditions - reduced - EMGFT})-(\ref{eq:Electric cov conditions - reduced - EMGFT}) reduce to
\begin{equation}
    \frac{\partial\{\tilde{\bar{F}}_{0x},\tilde{H}[\epsilon^0]\}}{\partial (\epsilon^0)'} = \frac{\partial\{\tilde{\bar{F}}_{0x},\tilde{H}[\epsilon^0]\}}{\partial (\epsilon^0)''} = 0
\end{equation}
for the U(1) Maxwell field,
\begin{eqnarray}\label{eq:Cov cond W0x3 - EMG - spherical}
    \frac{\partial\{\tilde{\bar{{\cal W}}}_{0x}^3,\tilde{H}[\epsilon^0]\}}{\partial (\epsilon^0)'} \!\!&=&\!\! \frac{\partial\{\tilde{\bar{{\cal W}}}_{0x}^3,\tilde{H}[\epsilon^0]\}}{\partial (\epsilon^0)''} = 0\,,\\
    \frac{\partial\{\tilde{\bar{{\cal W}}}_{0\vartheta}^i,\tilde{H}[\epsilon^0]\}}{\partial(\epsilon^0)'} \!\!&=&\!\! \tilde{q}^{xx} \tilde{\bar{{\cal W}}}_{x\vartheta}^i
    \,,\label{eq:W0thi - EMG}\\
    \frac{\partial\{\tilde{\bar{{\cal W}}}_{0\vartheta}^i,\tilde{H}[\epsilon^0]\}}{\partial(\epsilon^0)''} \!\!&=&\!\! 0
\end{eqnarray}
for the electric components of the SU(2) Yang--Mills strength tensor field,
\begin{eqnarray}\label{eq:Magnetic cov cond - EMGFT - spherical}
    \tilde{\bar{{\cal W}}}_{0 \vartheta}^i \!\!&=&\!\! \frac{\partial\tilde{H}}{\partial P_i}\,,\\
    \tilde{\bar{{\cal W}}}^1_{0\varphi} \!\!&=&\!\! - \sin\vartheta \tilde{\bar{{\cal W}}}^2_{0\vartheta}
    \qquad,\quad
    \tilde{\bar{{\cal W}}}^2_{0\varphi} = \sin\vartheta \tilde{\bar{{\cal W}}}^1_{0\vartheta}\,,
\end{eqnarray}
as well as (\ref{eq:W0x3 - EMGFT}) for the magnetic components, and
\begin{eqnarray}\label{eq:Vector potential cov conditions - reduced - EMGFT - spherical - 1}
    \frac{\partial\tilde{H}}{\partial (P^x)'}\bigg|_{\rm O.S.}
    =  \frac{\partial\tilde{H}}{\partial (P^x)''}\bigg|_{\rm O.S.} = 0\,,\\
    \frac{\partial\tilde{H}}{\partial P_i'}\bigg|_{\rm O.S.}
    =  \frac{\partial\tilde{H}}{\partial P_i''}\bigg|_{\rm O.S.} = 0\,,\label{eq:Vector potential cov conditions - reduced - EMGFT - spherical - 2}
\end{eqnarray}
for the SU(2) vector potential.

Similar to what we found in the analysis of the full four-dimensional theory, the magnetic covariance condition (\ref{eq:Magnetic cov cond - EMGFT - spherical}) can be used as the definition the angular components of the emergent SU(2) electric field despite them being absent from the constraint algebra as structure functions.

As usual in canonical theories, it is possible to use canonical transformations that preserve the form of the vector and Gauss constraints but otherwise change the form of other phase-space functions.
In doing so, one may obtain a Hamiltonian constraint and structure functions that look very different from their expressions prior to the transformations, but both in fact describe the same physical system.
A systematic procedure to factor out these canonical transformations was presented in \cite{EMGCov,EMGscalar,EMGEM}, where the first step is to map the modified angular component 
$\tilde{q}_{\vartheta \vartheta} ({\cal E}^x)$ to
its classical form $\tilde{q}_{\vartheta \vartheta} \to {\cal E}^x$, as well as to map other quantities related to Dirac observables (discussed below) closer to their respective classical forms; the remaining steps fix the canonical transformations in order to simplify the differential equations resulting from imposing the covariance conditions.

\subsubsection{Hamiltonian constraint and structure functions}

The implementation of anomaly freedom and the covariance conditions, as well as the existence of the symmetry generators ---corresponding to the mass and charge observables characteristic of the spherically symmetric system with no degrees of freedom---was performed in \cite{EMGEM} starting with the most general constraint ansatz involving up to second-order derivatives of the phase-space variables for the Einstein--Maxwell system, and previously for the vacuum system in \cite{EMGCov}.
Here, we generalize the results to include SU(2)-Yang--Mills degrees of freedom (the detailed derivation imposing the above covariance conditions can be found in the appendix), and couple the system to neutral scalar matter \cite{EMGscalar} and to a neutral perfect fluid \cite{EMGPF}.
The Hamiltonian constraint can be written as $\tilde{H}=\tilde{H}^{\rm EMYM}+\tilde{H}^\varphi+\tilde{H}^{\rm PF}$ with the Einstein--Maxwell--Yang--Mills contribution given by
\begin{widetext}
\begin{eqnarray}
    \label{eq:Hamiltonian constraint - final - EMYM - non-periodic}
    \tilde{H}^{\rm EMYM}
    \!\!&=&\!\! - \sqrt{{\cal E}^x} \frac{\chi}{2} \bigg[ {\cal E}^\varphi \bigg( \frac{\alpha_0}{{\cal E}^x}
    + 2 \frac{\sin^2 \left(\lambda K_\varphi\right)}{\lambda^2}\frac{\partial c_{f}}{\partial {\cal E}^x}
    + 4 \frac{\sin \left(2 \lambda K_\varphi\right)}{2 \lambda} \frac{1}{\lambda} \frac{\partial \left(\lambda q\right)}{\partial {\cal E}^x}
    \\
    \!\!&&\!\!
    \qquad\qquad\qquad
    + \left(\frac{\alpha_2}{{\cal E}^x} - 2 \frac{\partial \ln \lambda^2}{\partial {\cal E}^x}\right) \left( c_f \frac{\sin^2 \left(\lambda K_\varphi\right)}{\lambda^2}
    + 2 q \frac{\sin \left(2 \lambda K_\varphi\right)}{2 \lambda} \right)
    \nonumber\\
    \!\!&&\!\!
    \qquad\qquad\qquad
    + \left(4 \frac{K_x}{{\cal E}^\varphi} + \frac{\partial \ln \lambda^2}{\partial {\cal E}^x} 2 K_\varphi \right) \left(c_f \frac{\sin (2 \lambda K_\varphi)}{2 \lambda}
    + q \cos(2 \lambda K_\varphi)\right)
    \bigg)
    \nonumber\\
    \!\!&&\!\!
    \qquad\qquad
    + \frac{(({\cal E}^x)')^2}{{\cal E}^\varphi} \bigg(
    - \frac{\alpha_2}{4 {\cal E}^x} \cos^2 \left( \lambda K_\varphi \right)
    + \lambda^2 \left(\frac{K_x}{{\cal E}^\varphi}+\frac{K_\varphi}{2} \frac{\partial \ln \lambda^2}{\partial {\cal E}^x} \right)\frac{\sin \left( 2 \lambda K_\varphi \right)}{2 \lambda} \bigg)
    \nonumber\\
    \!\!&&\!\!
    \qquad\qquad
    + \left(\frac{({\cal E}^x)' ({\cal E}^\varphi)'}{({\cal E}^\varphi)^2}
    - \frac{({\cal E}^x)''}{{\cal E}^\varphi}\right) \cos^2 \left( \lambda K_\varphi \right)
    \nonumber\\
    \!\!&&\!\!
    \qquad\qquad
    - \frac{(P_1)^2+(P_2)^2}{{\cal E}^\varphi} \frac{1}{1+\theta^2 \alpha_3} \left(c_f+\theta^2 \alpha_4 - \lambda^2 \left( c_f \frac{\sin^2 \left(\lambda K_\varphi\right)}{\lambda^2}
    + 2 q \frac{\sin \left(2 \lambda K_\varphi\right)}{2 \lambda} \right)\right)
    \nonumber\\
    \!\!&&\!\!
    \qquad\qquad
    - (1+\theta^2)\frac{({\not\!B}_1^\varphi)^2+({\not\!B}^\varphi_2)^2}{{\cal E}^\varphi} \alpha_3
    - \frac{2\theta}{{\cal E}^\varphi} \left(P_1{\not\!B}^\varphi_2-P_2{\not\!B}^\varphi_1\right) \alpha_4
    - \lambda^2 \frac{(({\cal E}^x)')^2}{{\cal E}^\varphi} \frac{(P_1)^2+(P_2)^2}{{\cal E}^\varphi} \frac{\cos^2(\lambda K_\varphi)}{2(1+\theta^2)\alpha_3}
    \bigg]
    \,,\nonumber
\end{eqnarray}
a scalar matter contribution is given by \cite{EMGscalar}
\begin{eqnarray}
    \tilde{H}^\varphi \!\!&=&\!\!
    - \chi \frac{\sqrt{{\cal E}^x}}{2} \Bigg[- \frac{P_\phi{}^2}{{\cal E}^\varphi} \frac{\alpha_3}{{\cal E}^x} \left( \left(c_f+\lambda^2 \left(\frac{({\cal E}^x)'}{2{\cal E}^\varphi}\right)^2\right) \cos^2 (\lambda K_\varphi)
    - 2 q \lambda^2 \frac{\sin (2 \lambda K_\varphi)}{2 \lambda} \right)
    - \frac{\left( \phi' + c_{h3} ({\cal E}^x)' \right)^2}{{\cal E}^\varphi} \frac{{\cal E}^x}{\alpha_3}
    - 2 {\cal E}^\varphi V
    \nonumber\\
    \!\!&&\!\!
    \qquad\qquad
    + 4 c_h P_\phi \left( \left(c_f+\lambda^2 \left(\frac{({\cal E}^x)'}{2{\cal E}^\varphi}\right)^2\right) \frac{\sin (2 \lambda K_\varphi)}{2 \lambda}
    + q \cos(2 \lambda K_\varphi) \right)
    \Bigg]
    + \chi^2 {\cal E}^x \sqrt{\tilde{q}_{xx}} V_q
    + ({\cal E}^\varphi)^2 \sqrt{\tilde{q}^{xx}} V^q
    \,,\qquad
    \label{eq:Hamiltonian constraint - scalar}
\end{eqnarray}
\end{widetext}
and the perfect fluid contribution by \cite{EMGPF}
\begin{eqnarray}
    \tilde{H}^{\rm PF} \!\!&=&\!\!
    \sqrt{ s P_T^2 + \tilde{q}^{x x} \left(P_X X' + P_T T'\right)^2}
    \nonumber\\
    \!\!&&\!\!
    - {\cal E}^x \sqrt{\tilde{q}_{x x}} P_q
    - \sqrt{{\cal E}^x} {\cal E}^\varphi P_0
    \,,
    \label{eq:Hamiltonian constraint - PF - spherical - EMGFT}
\end{eqnarray}
with $s=1$ for a timelike fluid and $s=0$ for a null fluid, and $\theta$ is the SU(2) topological constant ---the U(1) topological constant drops out in the spherically symmetric system because the U(1) magnetic field is trivial.

The parameters $\lambda$, $c_f$, $q$, and $c_h$ are undetermined functions of the triad ${\cal E}^x$; $\alpha_0$, and $\alpha_2$ are undetermined functions of ${\cal E}^x$, ${\cal P}^x$, $P^x$, and ${\not\!B}^x$; and $\chi$ can is an undetermined function of ${\cal E}^x$, ${\cal P}^x$, $P^x$, and ${\not\!B}^x$, and $\phi$.
The scalar matter terms $V$, $V_q$, $V^q$ are undetermined functions of both ${\cal E}^x$ and $\phi$, describing three possible potentials.
The fluid terms $P_0$ and $P_q$ are undetermined functions of ${\cal E}^x$ and $P_X/P_T$, describing two possible pressure functions.
In the absence of scalar matter, $c_h$ and the potentials must vanish;
in the absence of the perfect fluid, the pressure functions must vanish.
The classical expressions of the constraints can be recovered in several ways, but the most straightforward is given by the limit $\lambda,q,c_h,V_q,V^q,P_q\to0$, $\chi,c_f,\alpha_2,\alpha_3,\alpha_4 \to 1$ and $\alpha_0\to 1-\Lambda {\cal E}^x-\left(({\cal P}^x)^2+(P^x+\theta {\not\!B}^x)^2+({\not\!B}^x)^2\right)/{\cal E}^x$, as well as the dependence reduction of $V(\phi,{\cal E}^x)\to V(\phi)$ and $P_0(P_X/P_T,{\cal E}^x)\to P_0(P_X/P_T)$.

The gravitational structure function associated to the modified constraint $\tilde{H}$ is given by
\begin{widetext}
\begin{equation}
    \tilde{q}^{x x} =
    \left(
    \left( c_f
    + \lambda^2 \left( \frac{({\cal E}^x)'}{2 {\cal E}^\varphi} \right)^2
    \right)
    \cos^2 \left( \lambda K_\varphi \right)
    - 2 q \lambda^2 \frac{\sin (2 \lambda K_\varphi)}{2 \lambda}
    \right) \chi^2
    \frac{{\cal E}^x}{({\cal E}^\varphi)^2}
    \,.
    \label{eq:Structure function - final}
\end{equation}
\end{widetext}
The nontrivial components of the strength tensor are given by equations (\ref{eq:F0x - EMGFT}) for $\tilde{\bar{F}}_{0x}$ and (\ref{eq:W0x3 - EMGFT}) for $\tilde{\bar{{\cal W}}}_{0x}^3$, which can be rather lengthy expressions depending on the dependence of the modification functions on the radial momenta, while the angular, electric components of the SU(2) strength tensor are given by
\begin{widetext}
\begin{eqnarray}\label{eq:Magnetic cov cond - EMGFT - spherical - explicit}
    \tilde{\bar{{\cal W}}}_{0 \vartheta}^1 \!\!&=&\!\! \chi \frac{\sqrt{{\cal E}^x}}{{\cal E}^\varphi} \left(\frac{P_1}{\alpha_3 \left(1+\theta^2\right)} \left(\left(c_f + \lambda^2 \left(\frac{(E^x)'}{2{\cal E}^\varphi}\right)^2\right)\cos^2\left(\lambda K_{\varphi}\right)
    - 2 \lambda^2 q \frac{\sin \left(2 \lambda K_{\varphi}\right)}{2 \lambda}+ \theta^2 \alpha_4^2\right)+ \theta \alpha_4 {\not\!B}_2^{\varphi}\right)\,,\\
    \tilde{\bar{{\cal W}}}_{0 \vartheta}^2 \!\!&=&\!\! \chi \frac{\sqrt{{\cal E}^x}}{{\cal E}^\varphi} \left(\frac{P_2}{\alpha_3 \left(1+\theta^2\right)} \left(\left(c_f + \lambda^2 \left(\frac{(E^x)'}{2{\cal E}^\varphi}\right)^2\right)\cos^2\left(\lambda K_{\varphi}\right)
    - 2 \lambda^2 q \frac{\sin \left(2 \lambda K_{\varphi}\right)}{2 \lambda}+ \theta^2 \alpha_4^2\right)
    - \theta \alpha_4 {\not\!B}_1^{\varphi}\right)\,,
\end{eqnarray}
\end{widetext}
and
\begin{equation}
    \tilde{\bar{{\cal W}}}^1_{0\varphi} = - \sin\vartheta \tilde{\bar{{\cal W}}}^2_{0\vartheta}
    \qquad,\quad
    \tilde{\bar{{\cal W}}}^2_{0\varphi} = \sin\vartheta \tilde{\bar{{\cal W}}}^1_{0\vartheta}\,,
\end{equation}
and the magnetic components are given by their classical expressions (\ref{eq:W - magnetic components - spherical}).

The system retains the Dirac observables (\ref{eq:YM observable - spherical}) and (\ref{eq:Local charge observable - U(1)}).
The existence of the nonlocal observable (\ref{eq:YM observable - spherical}) implies a conserved densitized current $\tilde{J}^\mu_{\rm SU(2)}$ with components
\begin{eqnarray}
    \tilde{J}^t_{\rm SU(2)} \!\!&=&\!\! - \mathfrak{G}_{\rm SU(2)}
    \,,\\
    \tilde{J}^a_{\rm SU(2)} \!\!&=&\!\!
    N \chi \frac{\sqrt{{\cal E}^x}}{{\cal E}^\varphi} \bigg[
    (1+\theta^2) \alpha_3 \left(W_1B^\varphi_2-W_2 B^\varphi_1\right)
    \\
    \!\!&&\!\!\qquad\quad
    + \theta \alpha_4 \left(P_1W_1+P_2W_2\right)
    \bigg]
    - \mathfrak{G}_{\rm SU(2)} N^x\,.\nonumber 
\end{eqnarray}

In the free-scalar-field case, when $V=V_q=V^q=0$, the system has the Dirac observable
\begin{equation}
    G_\phi [\beta] = \int {\rm d}^3 x \beta P_\phi
    \,,
    \label{eq:Scalar field symmetry generator - DF}
\end{equation}
where $\beta$ is a constant.
The associated conserved scalar-matter-current components $J_\phi^\mu$ are given by
\begin{eqnarray}
    J^t_\phi \!\!&=&\!\!P_\phi
    \,,\\
    J^x_\phi \!\!&=&\!\! N \frac{\chi}{\alpha_3} \frac{({\cal E}^x)^{3/2}}{{\cal E}^\varphi} \left( \phi'
    + c_{h3} ({\cal E}^x)' \right)
    - P_\phi N^x
    \,.
    \label{eq:Conserved matter current - DF - roots}
\end{eqnarray}

In the absence of local degrees of freedom --- setting the scalar matter and the perfect fluid variables to zero as well as $P_1=P_2=W_1=W_2=0$ ---the system has the following additional local Dirac observables,
\begin{widetext}
\begin{eqnarray}\label{eq:Mass observable}
    \mathcal{M}
    \!\!&=&\!\!
    d_0
    + \frac{d_2}{4} \int {\rm d} {\cal E}^x \left[\frac{\alpha_0}{{\cal E}^x} \exp \int {\rm d} {\cal E}^x \ \frac{\alpha_2}{2 {\cal E}^x}\right]
    \nonumber\\
    \!\!&&\!\!
    + \frac{d_2}{2} \left(\exp \int {\rm d} {\cal E}^x \frac{\alpha_2}{2 {\cal E}^x}\right)
    \left[c_f \frac{\sin^2\left(\lambda K_{\varphi}\right)}{\lambda^2}
    - \cos^2 (\lambda K_\varphi) \left(\frac{({\cal E}^x)'}{2 {\cal E}^\varphi}\right)^2
    + 2 q \frac{\sin \left(2 \lambda  K_{\varphi}\right)}{2 \lambda}
    \right]
    \,,\nonumber
\end{eqnarray}
\end{widetext}
associated to the mass of the system and
\begin{equation}\label{eq:Local charge observable - SU(2)}
    {\cal Q}_{\rm SU(2)}^{\rm E} = - \left(P^x+\theta{\not\!B}^x\right) =: - \tilde{P}^x
    \quad,\quad
    {\cal Q}_{\rm SU(2)}^{\rm M} = - {\not\!B}^x\,,
\end{equation}
associated to the electric and magnetic SU(2)-charges of the system, respectively.
(In this case, the magnetic charge is a constant ${\cal Q}_{\rm SU(2)}^{\rm M}=g/2$.)
In the expression above, $d_0$ and $d_2$ are undetermined functions of the electric momentum ${\cal P}^x$, with the classical limit given by $d_0\to0$ and $d_2 \to 1$.

\subsection{Dynamics}

\subsubsection{Nonsingular SU(2)$\times$U(1)-charged black hole}

Isolating the two most interesting modification functions $\lambda({\cal E}^x)$ and $\alpha_0({\cal E}^x,{\cal P}^x)$, by setting all the rest to their classical values, except for a constant $\chi=\chi_0$, and $\alpha_0=1-\Lambda {\cal E}^x-(\alpha_{\rm U(1)}+\alpha_{\rm SU(2)})/{\cal E}^x$, with arbitrary functions $\alpha_{\rm U(1)}({\cal P}^x)$ and $\alpha_{\rm SU(2)}(P^x,{\not\!B}^x)$, results in expressions for the emergent line element and the strength tensor that still deviate from their classical relations to the phase space.
In the absence of scalar matter and the perfect fluid, a static solution exists in Schwarzschild coordinates, defined by the canonical partial-gauge fixing
\begin{equation}\label{eq:Schwarzschild gauge}
    N^x=0\quad,\quad
    E^x=x^2\,,
\end{equation}
if the SU(2) pairs $(A_1,P_1)$ and $(A_2,P_2)$ vanish.
The resulting line element is given by \cite{EMGEM}
\begin{equation}
    {\rm d}s^2
    = - f(x) {\rm d}t^2 +\frac{{\rm d}x^{2}}{h(x)f(x)}
    + x^2{\rm d}\Omega^2\,,
    \label{eq:Line element - Schwarzschild}
\end{equation}
where
\begin{eqnarray}\label{eq:Line element - Schwarzschild - f}
    f(x)\!\!&=&\!\! 1-\frac{2 M}{x} - \frac{\Lambda x^2}{3} + \frac{\alpha_{{\rm U(1)}\times{\rm SU(2)}}}{x^2}
    \,,\\
    h(x)\!\!&=&\!\!\chi_0^2\left(1+\lambda(x)^2f(x)\right)\,.\label{eq:Line element - Schwarzschild - h}
\end{eqnarray}
Here, $M$ is the value of the mass observable (\ref{eq:Mass observable}) evaluated in the solution, and $\alpha_{{\rm U(1)}\times{\rm SU(2)}}=\alpha_{\rm U(1)}+\alpha_{\rm SU(2)}$.
The expressions (\ref{eq:Line element - Schwarzschild - f}) and (\ref{eq:Line element - Schwarzschild - h}) follow from inverting (\ref{eq:Mass observable}) for (\ref{eq:Structure function - final}) in the dynamical solution; hence, (\ref{eq:Line element - Schwarzschild - f}) would require a nontrivial integration of $\alpha_{{\rm U(1)}\times{\rm SU(2)}}$ in the variable ${\cal E}^x$ if it depends on the latter.

In addition to the usual horizon at $x=2M$, the line element (\ref{eq:Line element - Schwarzschild}) diverges at the coordinate $x_\lambda^{(i)}$ solving the equation $h(x_\lambda^{(i)})=0$, with the upper index labeling possible different solutions.
As shown in \cite{ELBH,EMGEM}, these coordinate singularities, if they exist, lie beyond the black hole and cosmological horizons, in the homogeneous regions ---which can be described by the same line element (\ref{eq:Line element - Schwarzschild}) upon relabeling $t\to x_{\rm h}$ and $x\to t_{\rm h}$ with the new radial and time coordinates $x_{\rm h}$ and $t_{\rm h}$, respectively ---, and correspond to reflection-symmetry surfaces ($\lambda K_\varphi=\pm\pi/2$ in phase space) beyond which the spacetime seems to repeat itself ---the two regions, however, are in fact different because observers can cross the reflection-symmetry surface with one region in their future and the other left behind in their past.
Therefore, there are only two possible radii $x_\lambda^-$ and $x_\lambda^+$, which describe a minimum and maximum radius, respectively.
If $\lambda(x)$ is a decreasing function, then no maximum radius develops.
Furthermore, the spacetime is nonsingular at the minimum radius only if the total charge function $\alpha_{{\rm U(1)}\times{\rm SU(2)}}=\alpha_{\rm U(1)}+\alpha_{\rm SU(2)}$ is below an upper limit determined by the mass and the specific details of $\lambda(x)$ \cite{EMGEM}.
One function of interest is $\lambda=\sqrt{\Delta/{\cal E}^x}=\sqrt{\Delta}/x$ with positive constant $\Delta$ and $\chi_0=1$ because it implies an approximately classical geometry at large scales if $\Delta$ is small.
In this case, the upper limit for the charge function is given by
\begin{equation}\label{eq:Critical EM holonomy parameter}
    \alpha_{{\rm U(1)}\times{\rm SU(2)}}^{\rm (max)} = \frac{\frac{1}{4} (3-\Delta\Lambda)^{-1} \left[\Delta - \Sigma^{1/3}\right]^2 - 9 m_{\rm Pl}^2 \Delta}{16 \Sigma^{2/3}}
    \,,
\end{equation}
with
\begin{eqnarray}
    \!\!&&\!\!\!\!\Sigma = \Delta \bigg[
    - \Delta^2
    - 90 (3-\Delta\Lambda) \Delta m_{\rm Pl}^2 
    \\
    \!\!&&\!\!\!\!
    + (3-\Delta\Lambda)^2\left( 6 m_{\rm Pl} \left(9 m_{\rm Pl}^2 + \frac{2 \Delta}{3-\Delta\Lambda}\right)^{3/2} + 162 m_{\rm Pl}^4\right)
    \bigg]
    ,\nonumber
\end{eqnarray}
and the Planck mass $m_{\rm Pl}$ has been chosen as a limiting lower value of the black hole's mass $M\ge m_{\rm Pl}$, since any sub-Planckian mass would require the quantization of the theory.
The correct upper bound of the charge function can be realized by an appropriate choice of $\alpha_{{\rm U(1)}\times{\rm SU(2)}}$.
In particular, an effective description of Wilson-like action modifications suggests the functions
\begin{eqnarray}\label{eq:Wilson action mod - spherical - U(1)}
    \alpha_{\rm U(1)} \!\!&=&\!\! \frac{\sin^2 (\mathfrak{a} {\cal P}^x)}{\mathfrak{a}^2}
    \,,\\
    \alpha_{\rm SU(2)} \!\!&=&\!\! \frac{\sin^2 (\mathfrak{b}_P \tilde{P}^x)}{\mathfrak{b}_P^2}
    + \frac{\sin^2 (\mathfrak{b}_B {\not\!B}^x)}{\mathfrak{b}_B^2}
    \,,\label{eq:Wilson action mod - spherical - SU(2)}
\end{eqnarray}
or some other variant in terms of trigonometric functions, where $\mathfrak{a}$, $\mathfrak{b}_P$, and $\mathfrak{b}_B$ play the role of holonomy parameters ---which are allowed to depend on the squared radius ${\cal E}^x$ and hence vanishing at large scales, but here we consider them constant for simplicity ---and mimic the role of $\lambda$ for the graviational variable $K_\varphi$, which can similarly be understood as a gravitational holonomy length.
With these two modifications, $\lambda$ and $\alpha_{{\rm U(1)}\times{\rm SU(2)}}$, the black hole is nonsingular if the holonomy parameters satisfy
\begin{equation}
    \frac{1}{\mathfrak{a}^2} + \frac{1}{\mathfrak{b}_P^2} + \frac{1}{\mathfrak{b}_B^2} \leq \alpha_{{\rm U(1)}\times{\rm SU(2)}}^{\rm (max)}\,,
\end{equation}
and both the spacetime and the strength tensor are necessarily emergent quantities with nontrivial relations to the phase-space ---modification functions $\alpha_{\rm U(1)}$ and $\alpha_{\rm SU(2)}$ other than (\ref{eq:Wilson action mod - spherical - U(1)}) and (\ref{eq:Wilson action mod - spherical - SU(2)}) can yield nonsingular geometries as long as their effects in the line-element term (\ref{eq:Line element - Schwarzschild - f}) are properly bounded.
Using (\ref{eq:Wilson action mod - spherical - U(1)}) and (\ref{eq:Wilson action mod - spherical - SU(2)}), the nontrivial components of the strength tensor fields in the dynamical solution are given by
\begin{equation}
    \tilde{F}_{0 x} = - \frac{\sin (2\mathfrak{a} Q_{\rm U(1)})/(2\mathfrak{a})}{4\pi x^2 f(x)}
    \,,
\end{equation}
for the U(1) field, and
\begin{eqnarray}
    \tilde{W}_{0 x}^3 \!\!&=&\!\! - \frac{\sin (2\mathfrak{b}_P Q_{\rm SU(2)})/(2\mathfrak{b}_P)}{x^2 \sqrt{4\pi f(x)}}
    \,,\\
    \tilde{W}_{\vartheta\varphi}^3 \!\!&=&\!\! - \sin\vartheta 
    \,,
\end{eqnarray}
for the SU(2) field, where $Q_{\rm U(1)}$, $Q_{\rm SU(2)}^{\rm E}$, $Q_{\rm SU(2)}^{\rm M}$ are the respective values of the local charge observables (\ref{eq:Local charge observable - U(1)}) and (\ref{eq:Local charge observable - SU(2)}) evaluated in the dynamical solution.
The connections can be made static
\begin{eqnarray}
    A_a{\rm d}x^a \!\!&=&\!\! A_x(x) {\rm d}x
    \,,\\
    W_a^i {\rm d}x^a\tau_i \!\!&=&\!\! \frac{W_x(x)}{\sqrt{4\pi}} \tau_3 {\rm d} x
    + \tau_3 \cos \vartheta {\rm d} \varphi
    \,,\label{eq:SU(2) connection - spherical}
\end{eqnarray}
by completing the gauge fixing with the potentials
\begin{eqnarray}
    A_t \!\!&=&\!\! - \frac{\sin(2\mathfrak{a} Q_{\rm U(1)})/(2\mathfrak{a})}{x}
    \,,\\
    W_t^3 \!\!&=&\!\! - \frac{\sin(2\mathfrak{b}_P Q_{\rm SU(2)})/(2\mathfrak{b}_P)}{\sqrt{4\pi} x}\,,
\end{eqnarray}
where the boundary conditions $\lim_{x\to\infty}A_t(x)=0$ and $\lim_{x\to\infty}W_t^4(x)=0$ were used to set additive constants to zero.
The functions $A_x(x)$ and $W_x(x)$ remain undetermined by the solutions to the equations of motion, implying that they are pure gauge: Indeed the U(1) and SU(2)-gauge transformations generated by ${\cal A}=-\int {\rm d}xA_x(x)$ and ${\cal B}=-\int {\rm d}xW_x(x)$, respectively, result in vanishing radial components of the connections,
\begin{equation}
    A_x \to 0\quad,\quad W_x\to0\,,
\end{equation}
while leaving the rest of the phase-space variables unchanged.

Further geometric and thermodynamic properties of the black hole solution, as well as the global spacetime structure, can be found in \cite{ELBH,EMGEM}.

\subsubsection{Homogenous collapse}

The interior region of the modified black hole is homogeneous.
Therefore, a good model for collapsing matter in the interior region is that of homogeneous spherically symmetric systems.
This case was studied in \cite{EMGscalar} for the free scalar field with zero charge and constant $\lambda$.
Here, we generalize those results to arbitrary $\lambda(x)$ for free scalar matter and dust, setting all other modification functions to their classical values, and neglecting all potential terms.
In anticipation to the specific gauge choice below, it is convenient to perform the canonical transformation
\begin{eqnarray}
    K_\varphi \!\!&\to&\!\! \frac{\bar{\lambda}}{\lambda} K_\varphi
    \quad,\quad
    {\cal E}^\varphi \to \frac{\bar{\lambda}}{\lambda} {\cal E}^\varphi
    \nonumber\\
    K_x \!\!&\to&\!\! K_x - {\cal E}^\varphi K_\varphi \frac{\partial \ln \lambda}{\partial {\cal E}^x}\,,
    \label{eq:Canonical transformations - periodic}
\end{eqnarray}
with a constant $\bar{\lambda}$ and all other phase-space variables unchanged ---this canonical transformation preserves the form of the vector and Gauss constraints, but changes the form of the Hamiltonian constraints and of the structure functions; being canonical, the physical system is preserved under this transformation, but the equations of motion turn out to be simpler in these phase-space coordinates mainly because the Hamiltonian constraint becomes fully periodic in the new $K_\varphi$ variable \cite{EMGCov}.
Setting all spatial derivatives to zero for a manifestly homogeneous solution, the gauge can be fixed by choosing a vanishing shift $N^x=0$ (to preserve manifest homogeneity under time evolution) and choosing the angular curvature as an internal time; in particular, we choose the time coordinate
\begin{equation}
    t_\varphi = - K_\varphi\,.
\end{equation}
The consistency condition $\dot{t}_\varphi=-\{K_\varphi,\tilde{H}[N]\}=1$ then fixes the lapse.
In this gauge, the reflection-symmetry surface, where the radius minimizes in the vacuum solution, occurs at the time $t_\varphi=\pi/(2\bar{\lambda})$.
The on-shell condition $H_x=0$ trivializes for homogeneous solutions, while $\tilde{H}=0$ is solved for $K_x$ in terms of the rest of the phase-space variables.

If the homogeneous spacetime is filled with dust only, then the lapse is given by
\begin{equation}
    N = - \frac{\bar{\lambda}}{\lambda} \frac{2 \sqrt{{\cal E}^x}}{\left(1+\left(1-4 {\cal E}^x \frac{\partial \ln \lambda}{\partial {\cal E}^x}\right) \frac{\sin^2(\bar{\lambda}t_\varphi)}{\lambda^2}\right)}
    \label{eq:K_phi EoM - Homogeneous - PF}
\end{equation}
and the relevant equations of motion by
\begin{widetext}
\begin{eqnarray}
    (\ln {\cal E}^x)^\bullet
    \!\!&=&\!\!
    - 4 \frac{\bar{\lambda}^2}{\lambda^2} \frac{\sin \left(2\bar{\lambda} t_{\varphi}\right)}{2\bar{\lambda}}
    \Bigg(
    1 + \frac{\bar{\lambda}^2}{\lambda^2} \left(1-4{\cal E}^x\frac{\partial \ln \lambda}{\partial {\cal E}^x}\right) \frac{\sin^2\left(\bar{\lambda} t_{\varphi}\right)}{\bar{\lambda}^2}
    \Bigg)^{-1}
    \label{eq:E^x EoM - Homogeneous - PF}
    \\
    \left(\ln\tilde{q}_{xx}\right)^\bullet
    \!\!&=&\!\!
    \frac{\frac{2\bar{\lambda}}{\tan(\bar{\lambda}t_\varphi)} \left(1 - \frac{\bar{\lambda}^2}{\lambda^2} \left(1-4{\cal E}^x\frac{\partial \ln \lambda}{\partial {\cal E}^x}\right) \frac{\sin^2\left(\bar{\lambda} t_{\varphi}\right)}{\bar{\lambda}^2}\right)
    - 2 \frac{\bar{\lambda}}{\lambda} \frac{2\bar{\lambda}}{\tan(2\bar{\lambda}t_\varphi)} \frac{P_T}{\sqrt{\tilde{q}_{xx}}}}{1 + \frac{\bar{\lambda}^2}{\lambda^2} \left(1-4{\cal E}^x\frac{\partial \ln \lambda}{\partial {\cal E}^x}\right) \frac{\sin^2\left(\bar{\lambda} t_{\varphi}\right)}{\bar{\lambda}^2}}
    - (\ln {\cal E}^x)^\bullet
    \,,
    \nonumber
    \label{eq:E^phi EoM - Homogeneous - PF}
\end{eqnarray}
\end{widetext}
with $P_T$ conserved and $\tilde{q}_{xx}=({\cal E}^\varphi)^2/\left({\cal E}^x\cos^2(\bar{\lambda}K_\varphi)\right)$.
While a minimum radius is again attained at the time $t_\varphi=\pi/(2\bar{\lambda})$, the $P_T$ term in (\ref{eq:E^phi EoM - Homogeneous - PF}) introduces a divergence in the evolution of the radial metric component.
Consequently, the homogeneous collapse solution of dust is singular at the minimum radius surface as can be concluded by inspection of the Ricci scalar that the emergent line-element implies.

If, on the other hand, the homogeneous spacetime is filled with free scalar matter only, then the lapse is given by
\begin{widetext}
\begin{equation}
    N = 2 \sqrt{{\cal E}^x} \frac{\bar{\lambda}}{\lambda} \left(1
    + \frac{\bar{\lambda}^2}{\lambda^2} \left(1-4 {\cal E}^x \frac{\partial\ln\lambda}{\partial {\cal E}^x}\right) \frac{\sin^2(\bar{\lambda}t_\varphi)}{\bar{\lambda}^2}
    + \frac{\bar{\lambda}^2}{\lambda^2} \frac{G_\phi^2}{{\cal E}^x \tilde{q}_{xx}}\right)^{-1}
    \,,
    \label{eq:K_phi time evolution - homogeneous - DF}
\end{equation}
where $G_\phi=P_\phi$ is a constant associated to the observable (\ref{eq:Scalar field symmetry generator - DF}), and the relevant equations of motion are given by
\begin{eqnarray}
    (\ln {\cal E}^x)^\bullet
    \!\!&=&\!\!
    - 4 \frac{\bar{\lambda}^2}{\lambda^2} \frac{\sin \left(2\bar{\lambda} t_{\varphi}\right)}{2\bar{\lambda}} \left(1 + \frac{\bar{\lambda}^2}{\lambda^2} \left(1-4 {\cal E}^x \frac{\partial\ln\lambda}{\partial {\cal E}^x}\right) \frac{\sin^2\left(\bar{\lambda} t_{\varphi}\right)}{\bar{\lambda}^2}
    + \frac{G_\phi^2}{{\cal E}^x \tilde{q}_{xx}}\right)^{-1}
    \,,
    \label{eq:E^x EoM - Homogeneous - DF}
    \\
    \left(\ln \tilde{q}_{xx}\right)^\bullet
    \!\!&=&\!\!
    \frac{2\bar{\lambda}}{\tan\left(\bar{\lambda} t_{\varphi}\right)}
    \frac{1
    - \frac{\bar{\lambda}^2}{\lambda^2} \left(1-4{\cal E}^x\frac{\partial\ln\lambda}{\partial {\cal E}^x}\right) \frac{\sin^2\left(\bar{\lambda} t_\varphi\right)}{\bar{\lambda}^2}
    - \frac{\bar{\lambda}^2}{\lambda^2} \frac{G_\phi^2}{{\cal E}^x \tilde{q}_{xx}}}{1 + \frac{\bar{\lambda}^2}{\lambda^2} \left(1-4{\cal E}^x\frac{\partial\ln\lambda}{\partial {\cal E}^x}\right)\frac{\sin^2\left(\bar{\lambda}  t_\varphi\right)}{\bar{\lambda}^2}
    + \frac{\bar{\lambda}^2}{\lambda^2} \frac{G_\phi^2}{{\cal E}^x \tilde{q}_{xx}}}
    - (\ln {\cal E}^x)^\bullet
    \,,\qquad
    \label{eq:E^phi EoM - Homogeneous - DF}
    \\
    \dot{\phi}
    \!\!&=&\!\!
    2 \frac{\bar{\lambda}^2}{\lambda^2} \cos(\bar{\lambda}t_\varphi) \frac{G_\phi}{\sqrt{{\cal E}^x \tilde{q}_{xx}}} \left(1
    + \frac{\bar{\lambda}^2}{\lambda^2} \left(1 - 4 {\cal E}^x \frac{\partial \ln \lambda}{\partial {\cal E}^x}\right) \frac{\sin^2(\bar{\lambda}t_\varphi)}{\bar{\lambda}^2}
    + \frac{\bar{\lambda}^2}{\lambda^2} \frac{G_\phi^2}{{\cal E}^x \tilde{q}_{xx}}\right)^{-1}
    \,.
    \label{eq:phi EoM - Homogeneous - DF}
\end{eqnarray}
\end{widetext}
These expressions are regular at the time $t_\varphi=\pi/(2\bar{\lambda})$, where the time derivatives of all metric components, as well as of the scalar field, vanish and correspond to a minimum of ${\cal E}^x$.
Furthermore, the right-hand-side of these expressions simply changes sign when crossing this surface (i.e., under the transformation $\delta\to-\delta$ in $t_\varphi=\pi/2+\delta$), implying that the two sides of the minimum radius surface are symmetric.
Therefore, this solution describes a nonsingular and symmetric bounce of collapsing scalar matter. (The divergence in the equations of motion at $t_\varphi=0$ and $t_\varphi=\pi/\bar{\lambda}$ must be understood as coordinate singularities which are present also in the vacuum, $G_\phi\to0$, where they correspond to the black hole's horizons \cite{ELBH,EMGscalar}. See \cite{EMGscalar} for the exact solution for constant $\lambda=\bar{\lambda}$.)

\section{Discussion}
\label{sec:Discussion}

We have extended the concept of emergent fields and the covariance conditions, first introduced for the gravitational field in \cite{EMGCov}, to Yang--Mills degrees of freedom.
This procedure results not only in an emergent spacetime metric field but also on emergent Yang--Mills force fields.
New theories of modified gravity, forces, and matter dynamics are available without introducing the typical additional degrees of freedom of higher-curvature terms or speculative new particles, and have been explicitly realized in several symmetry-reduced systems.
The modifications that are allowed after solving the covariance conditions are diverse and far from unique; however, their effects are computable and physical considerations can be used to further restrict them.
In particular, we have shown here that spherically symmetric geometries for SU(2)$\times$U(1)-charged black holes, as well as the homogeneous collapse of scalar matter, are robustly nonsingular for a certain class of modifications.

As expected, not all modifications resolve the singularities.
For instance, as shown here, the homogeneous collapse of dust leads to a singular geometry.
In fact, it has been shown that in Oppenheimer--Snyder \cite{EMGPF} and Lemaître--Tolman--Bondi collapse models \cite{EMGLTB} generically also develop singular emergent geometries.
Furthermore, other couplings of scalar matter are possible \cite{EMGscalar}; one such alternative realization is the minimal coupling, given by the Hamiltonian contribution $\tilde{H}^\phi=\frac{1}{2} \sqrt{\tilde{q}^{xx}} \left[P_\phi{}^2/{\cal E}^x + {\cal E}^x (\phi')^2\right] + \frac{1}{2} {\cal E}^x \sqrt{\tilde{q}_{xx}} V$ which generates equations of motion for the scalar field that are identical to the Klein--Gordon equation on the emergent spacetime $\tilde{g}_{\mu\nu}$.
The minimal coupling, however, leads to a singular homogeneous collapse \cite{EMGscalar}, and it is mathematically unsatisfactory from an effective field theory viewpoint because even the free-field case contains higher powers of derivative terms from $\sqrt{\tilde{q}^{xx}}$, which is not the case in the constraint contribution (\ref{eq:Hamiltonian constraint - scalar}) presented here and already derived in \cite{EMGscalar} if the modification functions $V_q$ and $V^q$ are discarded.
These observations can be used to discriminate between allowed modifications by their singularity-resolution status or other physical effects they generate.
For instance, the modified coupling (\ref{eq:Hamiltonian constraint - scalar}) is preferred and singled out by its nonsingular dynamical geometry.
Similarly, the generically singular nature of dust collapse leads us to not consider it as a fundamental form of matter ---a stance that is in agreement with the consensus and implies that its effects at extreme curvatures \emph{must} not be taken as conclusive phenomena in any theory.
Scalar matter, being a fundamental form of matter, implies more reliable results and turn out to include nonsingular collapsing geometries.

The different couplings of scalar matter imply not only differences in collapse solutions but also in several new dynamical phenomena, including the spectrum of quasinormal modes \cite{EMGscalarQNM}.
In particular, the nonsingular coupling, in contrast to the minimal coupling, predicts the instability of the quasinormal $s$-mode at a low critical mass $M_c$ whose value depends on the specifics of $\lambda(x)$, implying the mode's amplification rather than its decay if the black hole has a mass lower than this value.
If $\lambda=\sqrt{\Delta}/x$, this critical mass is given by $M_c\approx0.57\sqrt{\Delta}$.

Furthermore, Hawking evaporation can be studied by standard means using the background vacuum line-element (\ref{eq:Line element - Schwarzschild}) and the constraint (\ref{eq:Hamiltonian constraint - scalar}) to generate test-field dynamics, resulting in the usual spectrum of the frequency-dependent number of particles \cite{EMGHawking}
\begin{equation}\label{Hawkingdistribution2}
    \braket{N_{\omega}}=\frac{{\cal T}_{l}(\omega)}{{\rm exp}\left(\omega T_{\rm H}\right)-1}\,,
\end{equation}
with a black-body distribution of Hawking temperature $T_{\rm H}=1/(8\pi M)$, where new effects appear only in the greybody factor ${\cal T}_{l}(\omega)$.
The dominant $s$-wave contribution, ${\cal T}_0(\omega)$, can be treated using the standard low-frequency approximation, assuming the incoming scalar particle's energy is lower than the thermal energy, $ \omega\ll T_{\rm H}$ or $M\omega \ll 1$; for decreasing $\lambda=\sqrt{\Delta}/x$, this is given by
\begin{equation}
    {\cal T}_{0}(\omega)=\frac{16M^{2}\omega^{2}}{\left(1-4M^{2}\omega^{2}\right)^{2}+4M^{2}\omega^{2}\left(1+
    \frac{\Delta}{8 M^2}\right)^{2}}\,,
\end{equation}
and the energy emission rate by
\begin{equation}\label{eq:Energy emission}
    \frac{{\rm d}M}{{\rm d}t}=-\frac{1}{2\pi}\int_{0}^{\infty}{\rm d}\omega \frac{\omega \mathcal{T}_{0}(\omega)}{{\rm exp}(8\pi M\omega)-1}\,.
\end{equation}
This emission rate eventually slows down the evaporation process compared to the classical one: The emission rate initially increases similar to the classical case, but then sharply decreases at the mass $M_r\approx 0.15 \sqrt{\Delta}$, and zero mass is reached in an infinite amount of time \cite{EMGHawking}.
However, the black hole reaches the critical mass $M_c\approx0.57\Delta>M_r$ before this evaporation slowdown takes place and the dominant quasinormal $s$-mode becomes unstable, the black hole undergoes a rapid explosion taking over the Hawking emission rate, ending the evaporation process unambiguously.
If the $\lambda$ parameter is interpreted as an effective modification of quantum origin, then the constant $\Delta$ should be of the order of a Planck area and hence all these new effects occur near Planckian masses.
The negative net energy loss, together with the nonsingular status of the geometry, implies that the apparent horizon is timelike and matter can therefore escape the evaporating black hole, more easily so during its last stages where the net emission rate is high.

All the applications discussed above show profound new effects arising from short-range $\lambda$ modifications.
However, the emergent formulation allows for a great variety of modifications beyond the $\lambda$ function.
In particular, the $c_f$ modification function appears not only in the constraint (\ref{eq:Hamiltonian constraint - final - EMYM - non-periodic}) but also in the structure function (\ref{eq:Structure function - final}); it can therefore have important effects in modified spacetime solutions.
Indeed, this modification function was used in \cite{EMGMOND} to derive solutions with long-range gravitational effects that serve as relativistic realizations of modified Newtonian dynamics (MOND) ---and hence is an alternative solution to the dark matter problem without extra degrees of freedom ---with several other potential astronomical effects.

Moreover, examples of the emergent formulation are not restricted to spherical symmetry.
The modified vacuum constraint has also been derived in Gowdy systems \cite{EMGGowdy}, which have a local gravitational degree of freedom that makes it possible to study the effects of the modifications on the propagation of gravitational waves.
In particular, these Gowdy systems include modifications that are very similar to the $\lambda$ modifications of spherical symmetry and correspondingly lead to nonsingular bouncing geometries \cite{EMGGowdy}.

Additionally, the theory has recently been applied to Friedmann--Lemaître--Robertson--Walker (FLRW) cosmological models coupled to scalar matter and dust, including perturbative scalar degrees of freedom.
In this system, the covariance conditions are imposed on the perturbative inhomogeneities \cite{EMGCosmoK}, and the resulting constraints can be further restricted by other fundamental or physical considerations, including that they generate nonsingular background cosmologies, as well as dynamics for the massless degrees of freedom with their propagation speeds equal to the speed of light \cite{EMGCosmo,EMGCosmoD}.

We have therefore arrived at a promising post-Einstein--Yang--Mills theory of gravity and forces with several new physical effects.
An important next step in the theory is achieving the coupling of fermions, which are algebraically complicated to treat in canonical formulations, but are expected to follow covariance conditions that are conceptually similar to those of the Yang--Mills fields.
Furthermore, while it is encouraging that several realizations of the theory are possible in different symmetry-reduced settings, and with perturbative inhomogeneities, it remains to be seen whether modifications are possible in four-dimensional systems without symmetries, and whether such modifications are admitted under the (so far implicit) locality and continuum assumptions or if these should be relaxed.

Finally, if emergent field theory is used as the starting point of quantization, it would have important implications.
For instance, canonical quantization promotes the phase-space variables to operator-valued fields and replaces Poisson brackets for commutators.
While the commutation relations of the fundamental field operators would satisfy the Heisenberg algebra, the emergent fields related to spacetime geometry or forces generally would not because they must be understood as composite operators upon quantization.
This, in turn, implies that the spectrum of the emergent operators will have a nontrivial relation to the spectrum of the fundamental field operators; it is also possible that they do not even commute, in which case they do not share a complete set of simultaneous eigenstates.
Such as the concept emergent fields permits the formulation of covariant modifications to the classical theory of gravity, the possibility that composite operators, emerging from quantized gauge symmetries, could enable the quantization of gravity presents an intriguing new paradigm.

\section*{Acknowledgements}

The author thanks Martin Bojowald for useful discussions, in particular, regarding the spherical-symmetry reduction of SU(2) gauge fields.
This work was supported in part by NSF grant PHY-2206591.

\appendix

\section{Spherically symmetric SU(2) Yang--Mills phase space}

\subsection{Covariance conditions}

The nontrivial spacetime diffeomorphisms generated by $\xi^\mu\partial_\mu=\xi^t\partial_t+\xi^x\partial_x$ for a spherically symmetric SU(2) Yang--Mills 1-form $W_\mu^i$ and its strength tensor field ${\cal W}^i_{\mu\nu}$ respectively reduce to
\begin{eqnarray}\label{eq:Lie derivative 1-form - YM}
    \mathcal{L}_\xi W_t^3
    \!\!&=&\!\! \left(\epsilon^0 N^x-N\epsilon^x\right) {\cal W}_{0x}^3
    + \delta^{\rm SU(2)}_{\xi^\mu W_\mu^3} A_t^3
    \,,\\
    \mathcal{L}_\xi W_x^3 \!\!&=&\!\!
    \epsilon^0 {\cal W}_{0 x}^3
    + \delta^{\rm SU(2)}_{\xi^\nu W_\nu^3} W^3_x
    \,,\nonumber\\
    \mathcal{L}_\xi W_\vartheta^1 \!\!&=&\!\!
    \epsilon^0 {\cal W}_{0 \vartheta}^1
    + \epsilon^x {\cal W}_{x \vartheta}^1
    + \delta^{\rm SU(2)}_{\xi^\nu W_\nu^3} W^1_\vartheta
    \,,\nonumber\\
    \mathcal{L}_\xi W_\vartheta^2 \!\!&=&\!\!
    \epsilon^0 {\cal W}_{0 \vartheta}^2
    + \epsilon^x {\cal W}_{x \vartheta}^2
    + \delta^{\rm SU(2)}_{\xi^\nu W_\nu^3} W^2_\vartheta
    \,,\nonumber\\
    \mathcal{L}_\xi W_\varphi^1 \!\!&=&\!\!
    - \sin\vartheta \mathcal{L}_\xi W_\vartheta^2
    \quad,\quad
    \mathcal{L}_\xi W_\varphi^2 =
    \sin\vartheta \mathcal{L}_\xi W_\vartheta^1
    \,,\nonumber
\end{eqnarray}
and
\begin{eqnarray}\label{eq:Lie derivative magnetic components - YM}
    \mathcal{L}_\xi {\cal W}_{x \vartheta}^i \!\!&=&\!\!
    \frac{\epsilon^0}{N} \left(\dot{{\cal W}}_{x\vartheta}^i
    - \mathcal{L}_{\vec N} {\cal W}^i_{x\vartheta}
    - {\cal W}_{0 \vartheta}^i N'\right)
    \\
    \!\!&&\!\!
    + {\cal W}_{0 \vartheta}^i (\epsilon^0)'
    + \mathcal{L}_{\vec \epsilon} {\cal W}^i_{x\vartheta}
    \,,\nonumber\\
    \mathcal{L}_\xi {\cal W}_{x \varphi}^1 \!\!&=&\!\!
    - \sin\vartheta \mathcal{L}_\xi {\cal W}_{x \vartheta}^2
    \quad,\quad
    \mathcal{L}_\xi {\cal W}_{x \varphi}^2 =
    \sin\vartheta \mathcal{L}_\xi {\cal W}_{x \vartheta}^1
    \,,\nonumber\\
    \mathcal{L}_\xi {\cal W}_{\vartheta\varphi}^3 \!\!&=&\!\!
    \frac{\epsilon^0}{N} \left(\dot{{\cal W}}_{\vartheta\varphi}^3
    - \mathcal{L}_{\vec N} {\cal W}^3_{\vartheta\varphi}\right)
    + \mathcal{L}_{\vec \epsilon} {\cal W}^3_{\vartheta\varphi}
    \,,\nonumber\\
    \label{eq:Lie derivative electric components - YM}
    \mathcal{L}_\xi {\cal W}_{0 x}^3 \!\!&=&\!\!
    \frac{\epsilon^0}{N} \dot{{\cal W}}_{0 x}^3
    + \frac{1}{N} \left( N \epsilon^x
    - \epsilon^0 N^x \right) ({\cal W}_{0 x}^3)'
    \nonumber\\
    \!\!&&\!\!
    + \frac{1}{N} {\cal W}_{0 x}^3 \left( N (\epsilon^x)'
    - \epsilon^0 (N^x)' \right)
    \,,\nonumber\\
    \mathcal{L}_\xi {\cal W}_{0 \vartheta}^i \!\!&=&\!\!
    \frac{\epsilon^0}{N} \dot{{\cal W}}_{0 \vartheta}^i
    - \frac{1}{N} {\cal W}_{x\vartheta}^i q^{xx} \left(\epsilon^0 N' - N (\epsilon^0)' \right)
    \nonumber\\
    \!\!&&\!\!
    + \frac{1}{N} \left( N \epsilon^x
    - \epsilon^0 N^x \right) ({\cal W}_{0 \vartheta}^i)'
    \,,\nonumber\\
    \mathcal{L}_\xi {\cal W}_{0 \varphi}^1 \!\!&=&\!\!
    - \sin\vartheta \mathcal{L}_\xi {\cal W}_{0 \vartheta}^2
    \quad,\quad
    \mathcal{L}_\xi {\cal W}_{0 \varphi}^2 =
    \sin\vartheta \mathcal{L}_\xi {\cal W}_{0 \vartheta}^1
    \,.\nonumber
\end{eqnarray}

These Lie derivatives must be used in the right-hand-side of the covariance conditions (\ref{eq:Strength tensor covariance condition}) and (\ref{eq:YM 1-form covariance condition}).
Also, Poisson brackets must be used to evaluate the canonical gauge transformations of phase-space variables.
Because we have preserved the classical Gauss and vector constraints (and the density weights of the emergent fields are preserved due to the anomaly freedom conditions), only the normal transformations are relevant for the covariance conditions and given by
\begin{eqnarray}
    \{W_x^3,H[\epsilon^0]\}\!\!&=&\!\!
    \epsilon^0 {\cal W}_{0 x}^3
    \,,\label{eq:Ax3 - covcond}\\
    \{W_\vartheta^1,H[\epsilon^0]\}\!\!&=&\!\!
    \epsilon^0 {\cal W}_{0 \vartheta}^1
    \,,\label{eq:Ath1 - covcond}\\
    \{W_\vartheta^2,H[\epsilon^0]\}\!\!&=&\!\!
    \epsilon^0 {\cal W}_{0 \vartheta}^2
    \,.\label{eq:Ath2 - covcond}
\end{eqnarray}
for the vector potential,
\begin{eqnarray}\label{eq:F0x3 - covcond}
    \frac{\{{\cal W}_{0x}^3,H[\epsilon^0]\}}{\epsilon^0} \!\!&=&\!\!
    \frac{\{{\cal W}_{0x}^3,H[N]\}}{N}
    \,,\\
    \label{eq:F0thi - covcond}
    \frac{\{{\cal W}_{0\vartheta}^i,H[\epsilon^0]\}}{\epsilon^0}
    \!\!&-&\!\! {\cal W}_{x\vartheta}^i q^{xx} \frac{(\epsilon^0)'}{\epsilon^0}
    \\
    \!\!&&\!\!=
    \frac{\{{\cal W}_{0\vartheta}^i,H[N]\}}{N}
    - {\cal W}_{x\vartheta}^i q^{xx} \frac{N'}{N}
    \,,\nonumber
\end{eqnarray}
for the electric components of the strength tensor, and
\begin{eqnarray}
    \frac{\{{\cal W}_{\vartheta\varphi}^3,H[\epsilon^0]\}}{\epsilon^0} \!\!&=&\!\!
    \frac{\{{\cal W}_{\vartheta\varphi}^3,H[N]\}}{N}
    \,,\label{eq:Fthph3 - covcond}\\
    \frac{\{{\cal W}_{x\vartheta}^i,H[\epsilon^0]\}}{\epsilon^0} \!\!&-&\!\! {\cal W}_{0 \vartheta}^i \frac{(\epsilon^0)'}{\epsilon^0}
    \label{eq:Fxthi - covcond}\\
    \!\!&&\!\!=
    \frac{\{{\cal W}_{x\vartheta}^i,H[N]\}}{N}
    - {\cal W}_{0 \vartheta}^i \frac{N'}{N}
    \,,\nonumber
\end{eqnarray}
for the magnetic components, which are all understood to hold on shell.
The transformations of ${\cal W}^i_{x\varphi}$ and ${\cal W}^i_{0\varphi}$ are implied by those of ${\cal W}^i_{x\vartheta}$ and ${\cal W}^i_{0\vartheta}$, and the transformation of $W_\varphi^i$ is implied by that of $W_\vartheta^i$, hence their covariance conditions do not constitute independent equations.
The correct transformation of the Lagrange multiplier $W_t^3$ is implied by the anomaly-free constraint algebra.
These covariance conditions must hold for arbitrary gauge function $\epsilon^0$ and lapse $N$, implying the equations presented in Section \ref{sec:Covariance conditions - spherical} under the corresponding rescaling of factors of $4\pi$.

\subsection{Classical constraints}

Using the rescaled variables of Section~\ref{sec:Covariance conditions - spherical}, the only non-vanishing component of the SU(2) Gauss constraint in spherical symmetry is given by
\begin{equation}
    G^{\rm SU(2)}[W_t] = - \int{\rm d}x W_t \left[(P^x)'
    + {\not\!g} \left(P_2 W_1
    - P_1 W_2\right)\right]
    \,,
\end{equation}
and the SU(2) contribution to the vector constraint by
\begin{eqnarray}
    H^{\rm SU(2)}_x[N^x]
    \!\!&=&\!\! \int{\rm d}x N^x \big[P_1W_1'+P_2W_2'
    \\
    \!\!&&\!\!\qquad\qquad
    + {\not\!g} W_x (P_2W_1-P_1W_2)\big]
    \,,\nonumber
\end{eqnarray}
where we define the local expressions after integrating the angular coordinates.
Similarly, the contribution to the Hamiltonian constraint reduces to
\begin{widetext}

\begin{eqnarray}
    H^{\rm SU(2)}[N] \!\!&=&\!\!
    \int{\rm d}xN \frac{\sqrt{{\cal E}^x}}{2} \Bigg[\frac{{\cal E}^\varphi}{({\cal E}^x)^2} (P^x)^2 
    + \frac{(P_1)^2+(P_2)^2}{{\cal E}^\varphi}
    + (1+\theta^2) \left( {\cal E}^\varphi \frac{({\not\!B}^x)^2}{({\cal E}^x)^2}
    + \frac{({\not\!B}^\varphi_1)^2+({\not\!B}^\varphi_2)^2}{{\cal E}^\varphi}\right)
    \nonumber\\
    \!\!&&\!\!\qquad\qquad\qquad
    + 2 \theta \left( \frac{{\cal E}^\varphi}{({\cal E}^x)^2} P^x {\not\!B}^x
    + \frac{1}{{\cal E}^\varphi} \left(P_1 {\not\!B}^\varphi_2
    - P_2 {\not\!B}^\varphi_1\right)\right)\Bigg]
    \,,
\end{eqnarray}
\end{widetext}
and the symplectic term is given by
\begin{eqnarray}
    \int{\rm d}^3x P^a_i\dot{W}_a^i\!\!&=&\!\!
    \int{\rm d}x \left[P^x \dot{W}_x
    + P_1 \dot{W}_1
    + P_2 \dot{W}_2\right]
    \,.\qquad
\end{eqnarray}

\subsection{Constraint algebra}
The nontrivial SU(2)-invariant phase-space functions $f$ ---in the sense that $\{f,G^{\rm SU(2)}[{\cal B}]\}=0$ ---in spherical symmetry with no derivatives are given by
\begin{eqnarray}
    P^x
    \,&,&\,
    {\not\!B}^x
    \quad,\quad
    (P_1)^2+(P_2)^2\,,\nonumber\\
    {\not\!B}_+\!\!&=&\!\!{\not\!g}\left(P_1W_1 + P_2W_2\right)
    \,,\nonumber\\
    {\not\!B}_-\!\!&=&\!\!{\not\!g}\left(P_1W_2 - P_2W_1\right)
    \,,\label{eq:U(1)-inve - 0deriv}
\end{eqnarray}
and with first-order derivatives by
\begin{equation}
    ({\not\!B}^\varphi_1)^2+({\not\!B}^\varphi_2)^2
    \quad,\quad
    P_1 {\not\!B}^\varphi_2
    - P_2 {\not\!B}^\varphi_1
    \quad,\quad
    P_1 {\not\!B}^\varphi_1
    + P_2 {\not\!B}^\varphi_2
    \,,\label{eq:U(1)-inve - 1deriv}
\end{equation}
as well as by the derivative of the expressions in (\ref{eq:U(1)-inve - 0deriv}).

Because the constraints depend on the Yang--Mills phase-space variables only through the above SU(2)-invariant functions, it follows that
\begin{eqnarray}
    \{G^{\rm SU(2)}[W_t],G^{\rm SU(2)}[{\cal B}]\} \!\!&=& 0\,,\\
    \{H_x[N^x],G^{\rm SU(2)}[{\cal B}]\} \!\!&=&\!\! 0\,,\\
    \{H[N],G^{\rm SU(2)}[{\cal B}]\} \!\!&=&\!\! 0\,,
\end{eqnarray}
where $H_x=H_x^{\rm EH}+H_x^{\rm SU(2)}$ and $H=H^{\rm EH}+H^{\rm SU(2)}$ ---additional matter contributions may be incorporated accordingly.

Furthermore, using that the functional
\begin{eqnarray}\label{eq:Spatial diff generator SU(2) - spherical}
    D_x[\epsilon^x] \!\!&=&\!\! H_x[\epsilon^x] + G^{\rm SU(2)}[\epsilon^xW_x]
    \\
    \!\!&=&\!\! \int{\rm d}x N^x \big[{\cal E}^\varphi K_\varphi' - K_x ({\cal E}^x)'
    \nonumber\\
    \!\!&&\!\!\qquad\qquad
    + P_1 W_1'
    + P_2 W_2'
    - W_x (P^x)'\big]\nonumber
\end{eqnarray}
generates spatial diffeomorphisms of all phase-space variables, we can easily complete the constraint the algebra:
\begin{eqnarray}
    \!\!\{H_x[N^x],H_x[\epsilon^x]\} \!\!&=&\!\! - H_x[\epsilon^x(N^x)'-N^x(\epsilon^x)']\,,\qquad\\
    \!\!\{H[N],H_x[\epsilon^x]\} \!\!&=&\!\! - H[\epsilon^xN']
    + G^{\rm SU(2)} \left[N\epsilon^x\bar{F}_{0x}^3\right]\!,\qquad\\
    \!\!\{H[N],H[\epsilon^0]\} \!\!&=&\!\! - H_x[q^{xx}\left(\epsilon^0N'-N(\epsilon^0)'\right)]\,,\qquad
\end{eqnarray}
where
\begin{equation}
    \bar{F}_{0x}^3 = \frac{{\cal E}^\varphi}{({\cal E}^x)^{3/2}} \left(P^x+\theta {\not\!B}^x\right)\,.
\end{equation}

The canonical gauge transformation of the potential is then given by
\begin{eqnarray}
    \delta_{\epsilon,{\cal B}} W_t \!\!&=&\!\! - \dot{{\cal B}} + \bar{F}_{0x}^3\left(\epsilon^0 N^x- N\epsilon^x\right)\,,
\end{eqnarray}
in accordance with the covariance condition (\ref{eq:YM 1-form covariance condition}) using the symmetry-reduced Lie derivative (\ref{eq:Lie derivative 1-form - YM}).

\subsection{Modified constraint}
We choose our initial ansatz for the modified constraint contribution of the SU(2) Yang--Mills field based on the SU(2)-invariant functions (\ref{eq:U(1)-inve - 0deriv}) and (\ref{eq:U(1)-inve - 1deriv}) and the derivative order of the classical constraint:
\begin{widetext}
\begin{eqnarray}\label{eq:H SU(2) ansatz}
    \tilde{H}^{\rm SU(2)} \!\!&=&\!\!
    \chi \frac{\sqrt{{\cal E}^x}}{2} \Bigg[\frac{{\cal E}^\varphi}{({\cal E}^x)^2} a_0
    + \frac{(P_1)^2+(P_2)^2}{{\cal E}^\varphi} a_1
    + \frac{1+\theta^2}{{\cal E}^\varphi} ({\not\!B}^\varphi)^2 \left(a_3+\frac{(P_1)^2+(P_2)^2}{({\cal E}^\varphi)^2} b_3\right)
   \\
    \!\!&&\!\!\qquad\quad
    + \frac{2 \theta}{{\cal E}^\varphi} \left(P_1 {\not\!B}^\varphi_2
    - P_2 {\not\!B}^\varphi_1\right) \left(a_4+\frac{(P_1)^2+(P_2)^2}{({\cal E}^\varphi)^2} b_4\right)
    + \frac{(({\cal E}^x)')^2}{{\cal E}^\varphi} \left(c_1+\frac{(P_1)^2+(P_2)^2}{({\cal E}^\varphi)^2} c_2\right) \Bigg]\,.\nonumber
\end{eqnarray}
\end{widetext}
In this ansatz, the $a_I$, $b_I$, and $c_I$ coefficients are allowed to depend on ${\cal E}^x$, $P^x$, $K_\varphi$, and ${\not\!B}^x$, restricted so far only by compatibility with the classical limit and the requirement that they vanish in the vacuum limit ---here, $\chi$ is the same function of the vacuum constraint.
We also extend the phase-space dependence of the modification functions $\chi$, $c_f$, $q$ $\alpha_0$, and $\alpha_2$ of the gravitational contribution $\tilde{H}^{\rm EH}$ derived in \cite{EMGCov} to include $P^x$ and ${\not\!B}^x$.
The dependence on ${\cal E}^\varphi$ in the ansatz has been fixed for the constraint to remain a density of weight 1; no derivatives of the momenta were included due to the covariance conditions (\ref{eq:Vector potential cov conditions - reduced - EMGFT - spherical - 1}) and (\ref{eq:Vector potential cov conditions - reduced - EMGFT - spherical - 2}).

We now impose that, together with the vector and Gauss constraints, the Hamiltonian constraint $\tilde{H}=\tilde{H}^{\rm EH}+\tilde{H}^{\rm SU(2)}$ satisfies the Einstein--Yang--Mills algebra (\ref{eq:Hypersurface deformation algebra - spherical - modified}) with gravitational structure function (\ref{eq:Structure function - final}) ---only (\ref{eq:Hypersurface deformation algebra - HHa - spherical - modified}) and (\ref{eq:Hypersurface deformation algebra - HH - spherical - modified}) are nontrivial, and we disregard the U(1) Gauss constraint contribution in (\ref{eq:Hypersurface deformation algebra - HHa - spherical - modified}) in the following, which can be easily coupled as presented in the main text.
The bracket $\{\tilde{H}[N],G^{\rm SU(2)}[{\cal B}]\}=0$ is already realized by the ansatz (\ref{eq:H SU(2) ansatz}) because it is manifestly SU(2) invariant; the bracket (\ref{eq:Hypersurface deformation algebra - HHa - spherical - modified}) turns out to be satisfied too due to the ansatz having the correct density weight.
Using that the functional (\ref{eq:Spatial diff generator SU(2) - spherical}) generates spatial diffeomorphisms, it follows that the structure function in (\ref{eq:Hypersurface deformation algebra - HHa - spherical - modified}) is given by
\begin{equation}
    \tilde{\bar{{\cal W}}}_{0 x}^3 =
    \frac{\partial\tilde{H}}{\partial P^x}
    \,.\label{eq:F0x3 - EMG}
\end{equation}
Finally, the bracket (\ref{eq:Hypersurface deformation algebra - HH - spherical - modified}) is highly nontrivial and results in the following set of anomaly-freedom equations:
\begin{eqnarray}
    2 \lambda \sin(\lambda K_\varphi) c_2 + \cos(\lambda K_\varphi) \frac{\partial c_2}{\partial K_\varphi} \!\!&=&\!\! 0
    \,,\label{eq:AF - 1}\\
    2 \lambda \sin(\lambda K_\varphi) c_1 + \cos(\lambda K_\varphi) \frac{\partial c_1}{\partial K_\varphi} \!\!&=&\!\! 0
    \,,\\
    4 (1+\theta^2) a_3 c_2 - \lambda^2 \cos^2(\lambda K_\varphi) \!\!&=&\!\! 0
    \,,
\end{eqnarray}
\begin{eqnarray}
    b_3 \!\!&=&\!\! 0
    \,,\\
    b_4 \!\!&=&\!\! 0
    \,,\\
    \frac{\partial a_4}{\partial K_\varphi} \!\!&=&\!\! 0
    \,,
\end{eqnarray}
\begin{eqnarray}
    \frac{\partial c_f}{\partial P^x} \!\!&=&\!\! 0
    \,,\\
    \frac{\partial q}{\partial P^x} \!\!&=&\!\! 0\,,
\end{eqnarray}
\begin{eqnarray}
    4 ({\cal E}^x)^2 c_1 \left(c_f \sin(2\lambda K_\varphi)-2\lambda q\cos(2\lambda K_\varphi)\right) \!\!&&\!\!
    \\
    + \lambda \cos^2(\lambda K_\varphi) \frac{\partial a_0}{\partial K_\varphi} \!\!&=&\!\! 0
    \,,\nonumber\\
    c_f \lambda \sin(2\lambda K_\varphi) + 2 \lambda^2 q \cos(2\lambda K_\varphi) \!\!&&\!\!\\
    + (1+\theta^2) a_3 \frac{\partial a_1}{\partial K_\varphi} \!\!&=&\!\! 0
    \nonumber
    \,,\\
    \label{eq:AF - 9}
    2 (1+\theta^2) a_1 a_3 - 2 \theta^2 a_4^2 \!\!&&\!\!\\
    - 2 c_f \cos^2(\lambda K_\varphi)
    + 2 \lambda q \sin(2\lambda K_\varphi) \!\!&=&\!\! 0
    \,.\nonumber
\end{eqnarray}

The general solution to the set (\ref{eq:AF - 1})-(\ref{eq:AF - 9}) is given by
\begin{eqnarray}
    a_0 \!\!&=&\!\! \bar{a}_0 - 4 \beta\left(c_f \frac{\sin^2(\lambda K_\varphi)}{\lambda^2} + 2 q \frac{\sin(2\lambda K_\varphi)}{2\lambda}\right)
    \,,\\
    a_1 \!\!&=&\!\! \frac{1}{(1+\theta^2) \alpha_3} \bigg[c_f+\theta^2 \alpha_4^2
    \\
    \!\!&&\!\!\qquad\qquad
    - \lambda^2 \left(c_f\frac{\sin^2(\lambda K_\varphi)}{\lambda^2} + 2q\frac{\sin(2\lambda K_\varphi)}{2\lambda}\right)\bigg]
    \,,\nonumber\\
    a_3 \!\!&=&\!\! \alpha_3
    \quad,\quad
    a_4 = \alpha_4
    \,,\\
    b_3 \!\!&=&\!\!
    b_4 = 0
    \,,\\
    c_1 \!\!&=&\!\! \frac{\beta}{(E^x)^2} \cos^2(\lambda K_\varphi)
    \quad,\quad
    c_2 = \frac{\lambda^2 \cos^2(\lambda K_\varphi)}{4 (1+\theta^2) \alpha_3}
    \,,
\end{eqnarray}
where $\bar{a}_0$, $\alpha_3$, $\alpha_4$, and $\beta_1$ are undetermined functions of ${\cal E}^x$, $P^x$, and ${\not\!B}^x$, and the functions $c_f$ and $q$ cannot depend on $P^x$.
The functions $\bar{a}_0$ and $\beta$ can be absorbed into $\alpha_0$ and $\alpha_2$, respectively.
The resulting modified constraint ---coupled to electromagnetism ---is therefore given by (\ref{eq:Hamiltonian constraint - final - EMYM - non-periodic}).

The covariance conditions (\ref{eq:Spacetime cov condition - reduced - EMGFT - spherical xx}), (\ref{eq:Spacetime cov condition - reduced - EMGFT - spherical thth}), and (\ref{eq:Cov cond W0x3 - EMG - spherical})-(\ref{eq:Vector potential cov conditions - reduced - EMGFT - spherical - 2}) are all satisfied by the constraint (\ref{eq:Hamiltonian constraint - final - EMYM - non-periodic}) ---notice that some of these conditions had already been taken into account when formulating the ansatz (\ref{eq:H SU(2) ansatz}) and not the result of anomaly-freedom.


\begin{thebibliography}{10}

\bibitem{Lovelock1}
D. Lovelock, J. Math. Phys. {\bf 12}, 498--501 (1971).

\bibitem{Lovelock2}
D. Lovelock, J. Math. Phys. {\bf 13}, 874--876 (1972).

\bibitem{Weinberg}
S. Weinberg, {\em The Quantum Theory of Fields, Volume II: Modern Applications}, (Cambridge University Press, 1995).

\bibitem{ostrogradsky1}
M. Ostrogradsky, {\em Mémoire sur les équations différentielles relatives an probléme des isopérimétres}, (Imp. de l’Académie, 1848).

\bibitem{ostrogradsky2}
R.~P. Woodard, {\em The theorem of ostrogradsky}, (2015), arXiv:1506.02210.

\bibitem{woodard}
R.~P. Woodard, {\em Avoiding dark energy with 1/r modifications of gravity}, In {\em The invisible
universe: Dark matter and dark energy}, 403--433 (Springer, 2007).

\bibitem{ADM}
R. Arnowitt, S. Deser, and C.~W. Misner,  in {\em Gravitation: An Introduction to Current Research}, edited by L. Witten (Wiley, New York, 1962), reprinted in \cite{arnowitt2008republication}.

\bibitem{arnowitt2008republication}
R. Arnowitt, S. Deser, and C.~W. Misner, Gen.\ Rel.\ Grav. {\bf 40}, 1997 (2008).

\bibitem{hojman1976geometrodynamics}
S.~A. Hojman, K. Kucha\v{r}, and C. Teitelboim, Ann.\ Phys.\ (New York) {\bf 96},  88  (1976).

\bibitem{kuchar1974geometrodynamics}
K.~V. Kucha\v{r}, J.\ Math.\ Phys. {\bf 15}, 708 (1974).

\bibitem{Dirac}
P.~A.~M. Dirac, {\em The theory of gravitation in Hamiltonian form}, Proc. R. Soc. Lond. A, {\bf 246}, 333--343 (1958).

\bibitem{Katz}
J. Katz, {\em Les crochets de Poisson des contraintes du champ gravitationne}, C. R. Acad. Sci., Paris {\bf 254} 1386--7 (1962).

\bibitem{EMG}
M. Bojowald, and E.~I. Duque, Class.\ Quant.\ Grav. \textbf{42}, 095008 (2024), arXiv:2404.06375.

\bibitem{EMGCov}
M. Bojowald, and E.~I. Duque, Phys.\ Rev.\ D {\bf 108}, 084066, (2023), 	arXiv:2310.06798.

\bibitem{EMGEM}
E.~I. Duque, Phys.\ Rev.\ D {\bf 110}, 125006 (2024), 	arXiv:2407.14954.

\bibitem{EMGscalar}
M. Bojowald and E.~I. Duque, Phys.\ Rev.\ D 109, 084006, (2024) arxiv:2311.10693.

\bibitem{EMGPF}
E.~I. Duque, Phys.\ Rev.\ D {\bf 109}, 044014 (2024), 	arXiv:2311.08616.

\bibitem{Alonso_Bardaji_2022}
A. Alonso-Bardaj\'{\i}, D. Brizuela, and R. Vera, Phys.\ Lett.\ B {\bf 829}, 137075 (2022), arXiv:2112.12110.

\bibitem{ELBH}
I.~H. Belfaqih, M. Bojowald, S. Brahma, and E.~I. Duque,
\textit{Black holes in effective loop quantum gravity: Covariant holonomy modifications}, arXiv:2407.12087

\bibitem{EMGGowdy}
M. Bojowald and E.~I. Duque, Phys.\ Rev.\ D, {\bf 110}, 124001, (2024), arXiv:2407.13583.

\bibitem{EMGCosmoK}
M. Bojowald, M. Díaz, E.~I. Duque, \textit{Perturbative emergent modified gravity on cosmological backgrounds: Kinematics}, arXiv:2507.14358.

\bibitem{EMGCosmo}
M. Bojowald, M. Díaz, E.~I. Duque, \textit{Singularities in loop quantum cosmology}, arXiv:2507.08116.

\bibitem{EMGCosmoD}
M. Bojowald, M. Díaz, E.~I. Duque, \textit{Perturbative emergent modified gravity on cosmological backgrounds: Dynamics}, in preparation.

\bibitem{f(R)}
N. Deruelle, Y. Sendouda, and A. Youssef, Phys.\ Rev.\ D {\bf 80}, 084032 (2009), 	arXiv:0906.4983.

\bibitem{pons1997gauge}
J.~M. Pons, D.~C. Salisbury, and L.~C. Shepley, Phys.\ Rev.\ D {\bf 55}, 658 (1997), arXiv:gr-qc/9612037.

\bibitem{salisbury1983realization}
D.~C. Salisbury, and K. Sundermeyer, Phys.\ Rev.\ D {27}, 740 (1983).

\bibitem{bojowald2018effective}
M. Bojowald, S. Brahma, and D.~H. Yeom, Phys.\ Rev.\ D {\bf 98}, 046015 (2018), arXiv:1803.01119.

\bibitem{tHooft}
G. 't Hooft, Nucl. Phys. B {\bf 79}, 276 (1974).

\bibitem{Polyakov}
A.~M. Polyakov, JETP Lett. {\bf 20}, 194 (1974).

\bibitem{bojowald2000symmetry}
M. Bojowald, and H. Kastrup, Class.\ Quant.\ Grav. {\bf 17}, 3009 (2000), arXiv:hep-th/9907042.

\bibitem{bojowald2004spherically}
M. Bojowald, Class.\ Quant.\ Grav. {\bf 21}, 3733 (2004), arXiv:gr-qc/0407017.

\bibitem{EMGLTB}
M. Bojowald, E.~I. Duque, and D. Hartmann, Phys.\ Rev.\ D, {\bf 111}, 064002, (2025), arXiv:2412.18054.

\bibitem{EMGscalarQNM}
M. Bojowald, E.~I. Duque, and S. Shankaranarayanan, Phys.\ Rev.\ D, \textbf{111}, 024051, (2025) arXiv:2410.17501.

\bibitem{EMGHawking}
I.~H. Belfaqih, M. Bojowald, S. Brahma, and E.~I. Duque,
\textit{Hawking evaporation and the fate of black holes in loop quantum gravity}, arXiv:2504.11998.

\bibitem{EMGMOND}
M. Bojowald, and E.~I. Duque, Phys.\ Lett.\ B {\bf 847}, 138279 (2023), arXiv:2310.19894.

\end{thebibliography}
\end{document}